\documentclass[12pt]{iopart}

\usepackage{iopams}

\usepackage{graphicx,psfrag}
\usepackage{mathrsfs}
\usepackage{multirow}
\usepackage{accents}
\usepackage{comment}
\usepackage{ulem} \usepackage{multirow}
\usepackage{hyperref}
\usepackage{cite}
\usepackage{lipsum}

\newcommand{\be}{\begin{equation}}
\newcommand{\ee}{\end{equation}}
\newcommand{\bea}{\begin{eqnarray}}
\newcommand{\eea}{\end{eqnarray}}
\newcommand{\bel}{\begin{align}}
\newcommand{\eel}{\end{align}}

\def\lm{{\ell m}}

\def\non{\nonumber}

\def\e{{\rm e}}
\def\i{{\rm i}}
\def\d{{\rm d}}
\def\P{{\rm P}}

\def\Msun{{\rm M_{\odot}}}
\def\Mmax{{M_{\rm max}^{\rm TOV}}}
\def\Cmax{{C_{\rm max}^{\rm TOV}}}

\def\R14{{R_{1.4}^{\rm TOV}}}

\def\kt{\kappa^{\rm T}_2}

\def\GMc2{{\rm G M_{\odot} c^{-2}}}

\def\core{{\scshape CoRe}}

\def\spin{{\boldsymbol \chi}}

\newcommand{\THC}{\texttt{THC}}
\newcommand{\watpy}{\texttt{watpy}}
\newcommand{\BAM}{\texttt{BAM}}
\newcommand{\SGRID}{{\texttt{SGRID}}}
\newcommand{\Lorene}{{\texttt{Lorene}}}
\newcommand{\Cactus}{{\texttt{Cactus}}}

\begin{document}

\title[\core~DB R2]{Second release of the \core~database of binary neutron star merger
  waveforms}

\author{%
  Alejandra Gonzalez ${}^{1}$,
  Francesco Zappa ${}^{1}$, 
  Matteo Breschi ${}^{1}$, 
  Sebastiano Bernuzzi ${}^{1}$,
  David Radice ${}^{2,3,4}$,
  Ananya Adhikari ${}^{5}$, Alessandro Camilletti ${}^{6,7}$, Swami Vivekanandji Chaurasia ${}^{8}$, 
Georgios Doulis ${}^{1,9}$, Surendra Padamata ${}^{2,3}$, Alireza Rashti${}^{2,3}$, Maximiliano Ujevic ${}^{10}$,
  Bernd Br{\"u}gmann ${}^{1}$, William Cook ${}^{1}$, Tim Dietrich ${}^{11,12}$, Albino Perego ${}^{6,7}$, Amit Poudel ${}^{5}$, Wolfgang Tichy ${}^{5}$
}
\address{${}^{1}$Theoretisch-Physikalisches Institut, Friedrich-Schiller-Universit{\"a}t Jena, 07743, Jena, Germany}
\address{${}^{2}$Institute for Gravitation \& the Cosmos, The Pennsylvania State University, University Park PA 16802, USA}
\address{${}^{3}$Department of Physics, The Pennsylvania State University, University Park PA 16802, USA}
\address{${}^{4}$Department of Astronomy \& Astrophysics, The Pennsylvania State University, University Park PA 16802, USA}
\address{${}^{5}$Department of Physics, Florida Atlantic University, Boca Raton, FL 33431, USA}
\address{${}^{6}$Dipartimento di Fisica, Universit\'a di Trento, via Sommarive 14, 38123 Trento, Italy}
\address{${}^{7}$INFN-TIFPA, Trento Institute for Fundamental Physics and Applications, via Sommarive 14, 38123 Trento, Italy}
\address{${}^{8}$The Oskar Klein Centre, Department of Astronomy, Stockholm University, AlbaNova, SE-10691 Stockholm, Sweden}
\address{${}^{9}$Department of Physics, University of Athens, 15783, Athens, Greece}
\address{${}^{10}$Centro de Ci\^encias Naturais e Humanas, Universidade Federal do ABC, 09210-170, Santo Andr\'e, S\~ao Paulo, Brazil}
\address{${}^{11}$Institut f\"{u}r Physik und Astronomie, Universit\"{a}t Potsdam,
Haus 28, Karl-Liebknecht-Str.~24/25, 14476, Potsdam, Germany}
\address{${}^{12}$Max Planck Institute for Gravitational Physics (Albert Einstein Institute), Am Mühlenberg 1, Potsdam, Germany} 

\vspace{10pt}

\begin{indented}
\item[]\today{}
\end{indented}

\begin{abstract}
  We present the second data release of gravitational waveforms from
  binary neutron star merger simulations performed by the Computational
  Relativity (\core) collaboration. The current database consists of
  254 different binary neutron star configurations and a total of 590 
  individual numerical-relativity simulations using various grid resolutions. The
  released waveform data contain the strain and the
  Weyl curvature multipoles up to $\ell=m=4$. They 
  span a significant portion of the mass, mass-ratio,
  spin and eccentricity parameter space and include targeted
  configurations to the events GW170817 and GW190425.
  \core~simulations are performed with
  18 different equations of state, seven of which are finite
  temperature models, and three of which account for non-hadronic
  degrees of freedom. About half of the released data are computed
  with high-order hydrodynamics schemes for tens of orbits to merger; 
  the other half is computed with advanced microphysics.   We showcase a standard waveform
  error analysis and discuss the accuracy of the database in terms of
  faithfulness. We present ready-to-use fitting formulas for
  equation of state-insensitive relations at merger (e.g.\ merger frequency),
  luminosity peak, and post-merger spectrum. 
\end{abstract}

\section{Introduction}

The first observation of gravitational waves (GWs) from a binary neutron star (BNS) coalescence accompanied by electromagnetic (EM) signals marked a milestone in GW astronomy. Numerical 
relativity (NR) simulations are the main tool to explore the merger dynamics in the strong-field regime %%\cite{Shibata:1999wm,Shibata:2000jt}%%,Shibata:2002jb,Shibata:2003iw,Shibata:2003iy,Shibata:2003ga}
and aid the development of BNS gravitational waveforms that are necessary for GW detection and parameter estimation. The largest NR waveforms public catalogs contain data from thousands of 
binary black holes (BBHs) simulations, covering a significant portion of the mass-ratio and spin parameter space for quasi-circular 
mergers \cite{Mroue:2013xna,Chu:2015kft,Boyle:2019kee,Healy:2017psd,Healy:2019jyf,Healy:2020vre,Jani:2016wkt} and explore mergers from eccentric and generic 
orbits \cite{Healy:2017psd,Healy:2019jyf,Healy:2020vre}. Public waveforms from simulations of binaries with NSs are more limited, and include the \core~ database \cite{Dietrich:2018phi} 
(164 binaries at different resolutions for a total of 367), the SACRA-MPI \cite{Kiuchi:2017pte,Kiuchi:2019kzt} (46, total 276), the SXS (2, total 6)~\cite{Foucart:2018lhe,Boyle:2019kee}, among others~\cite{StellarCollapse:catalog,BNSwaveforms:catalog}. 
These waveforms are crucial for developing accurate inspiral-merger GW templates with tidal 
effects \cite{Baiotti:2011am,Bernuzzi:2012ci,Bernuzzi:2014owa,Hinderer:2016eia,Steinhoff:2016rfi,Bernuzzi:2016pie,Dietrich:2017aum,Kawaguchi:2018gvj,Nagar:2018zoe,Nagar:2018plt,Akcay:2018yyh,Isoyama:2020lls,Gamba:2021ydi} and postmerger emission \cite{Hotokezaka:2013iia,Bernuzzi:2015rla,Bauswein:2015yca,Clark:2015zxa,Easter:2018pqy,Tsang:2019esi,Breschi:2019srl,Easter:2020ifj,Soultanis:2021oia,Wijngaarden:2022sah,Whittaker:2022pkd,Breschi:2022xnc} with direct applications to equation of state (EOS) constraints \cite{Abbott:2017dke,Abbott:2018wiz,Abbott:2018exr,LIGOScientific:2019eut}. The NR simulations performed for these waveforms are also key to determine the properties of the remnants from the binary parameters and the input physics (EOS, mass, spins, etc.), e.g. \cite{Shibata:2005ss,Shibata:2006nm,Hotokezaka:2011dh,Bernuzzi:2013rza,Dietrich:2016hky,Dietrich:2016lyp,Dietrich:2017xqb,Koppel:2019pys,Baumgarte:1999cq,Hotokezaka:2013iia} (see also \cite{Radice:2020ddv,Bernuzzi:2020tgt} for recent reviews). Consequently, new and extended data releases are necessary to support research in the field of GW astronomy.

Here, we present a new release of the \core\ database that comprises 90 new physically distinct BNS configurations at multiple resolutions, for a total of 254 binaries and 590 simulations. 
The new release includes GW strains and Weyl multipoles information up to the $(\ell,m)=(4,4)$ mode. The new data were computed in simulations presented in Refs.~\cite{Radice:2018pdn,Perego:2019adq,Endrizzi:2019trv,Poudel:2020fte,Bernuzzi:2020txg,Nedora:2020pak,Chaurasia:2020ntk,Radice:2020ids,Dudi:2021abi,Prakash:2021wpz,Ujevic:2022qle,Doulis:2022vkx,Camilletti:2022jms} and include BNS waveforms consistent with the GW events GW170817~\cite{Perego:2017wtu,Endrizzi:2019trv,Bernuzzi:2020txg,Nedora:2020pak} and GW190425~\cite{Dudi:2018jzn,Camilletti:2022jms}.

The paper is organized as follows.
Sec.~\ref{sec:methods} summarizes the employed simulation techniques.
Sec.~\ref{sec:overview} describes the physics content of the database and the impact of the binary parameters on the waveforms.
Sec.~\ref{sec:acc} presents a full merger waveform error analysis for a case study, and gives an overview of the average accuracy of the data.
Sec.~\ref{sec:QUR} presents, as a first application of the database, ready-to-use EOS-insensitive fitting formulas for the GW frequency, amplitude and the peak luminosity that characterize the merger as well as analogous relations for the post-merger GW spectra.

The \core~ database is hosted on the public \texttt{gitlab} server
\begin{center}
\url{https://core-gitlfs.tpi.uni-jena.de/core_database}
\end{center}
Associated code repositories and resources can be accessed from
\begin{center}
\url{http://www.computational-relativity.org/}
\end{center}
In particular, we provide the python package \href{https://git.tpi.uni-jena.de/core/watpy}{\watpy} to ease the checkout of the data and perform standard waveform analyses.

\subsection{Notation}

NR data are computed in geometrized units $c=G=1$ and solar masses $\Msun = 1$; we use these units also in this paper unless explicitly indicated.
We recall that $GM_\odot/c^3\simeq4.925490947~\mu$s and $GM_\odot/c^2\simeq1.476625038$~km.
The binary mass is $M= m_1 +m_2$, where $m_{1,2}$ are the gravitational masses of the two stars.
The mass ratio is defined as $q = m_1 /m_2 \ge 1$, and the symmetric mass ratio is $\nu = m_1 m_2 / M^2\in[0,1/4]$, where $\nu=1/4$ corresponds to the equal-mass case, whereas $\nu\rightarrow 0$ for very unequal masses.
The dimensionless, mass-rescaled spin vectors are denoted with $\spin_i$ for $i=1,2$ and the spin components aligned with the orbital angular momentum
$\textbf{L}$ are labeled as $\chi_i = \spin_{i}\cdot \textbf{L} / |\textbf{L}|$.
The effective spin parameter $\chi_{\rm eff}$ is 
defined as the mass-weighted aligned spin,
\be
\label{eq:chieff}
\chi_{\rm eff} = \frac{m_1 \chi_{1}+m_2 \chi_{2}}{M}\,.
\ee
Similarly, one can define the spin parameter~\cite{Nagar:2019wds},
\be
\label{eq:shat}
\hat{S} = \left( \frac{m_1}{M} \right)^2 \chi_1 + \left( \frac{m_2}{M} \right)^2 \chi_2.
\ee
The quadrupolar tidal polarizability parameters are defined
as $\Lambda_{i}=({2}/{3})\,k_{2,i}\,C_i^{-5}$ for $i=1,2$~\cite{Damour:2009wj},
where $k_{2,i}$ and $C_i$ are respectively the $\ell=2$ gravito-electric Love numbers and the compactness of the $i$-th neutron star (NS). 
The tidal coupling constant is~\cite{Damour:2009wj}~\footnote{$\ell=2$ case of Equation 25 in~\cite{Damour:2009wj}. Here we use indices $1$ and $2$ to denote each star instead of $A$ and $B$. Note that $(1\leftrightarrow 2)$ indicates that the previous term with index $1$ is repeated with index $2$.}
\be
\label{eq:k2t}
\kt =3\nu\,\left[\left(\frac{m_1}{M}\right)^3 \Lambda_1 + (1\leftrightarrow 2)\right]\,,
\ee
that, similarly to the reduced tidal parameter~\cite{Favata:2013rwa}
\be
\tilde{\Lambda} = \frac{16}{13}\frac{(m_1 + 12m_2)m_1^4\Lambda_1}{M^5}+(1 \leftrightarrow 2)\,,
\ee
parametrizes the leading-order tidal contribution to the binary interaction potential and waveform phase (note that $\kt = \frac{3}{16}\tilde{\Lambda}$ for $q=1$.)

The radiated GW (polarizations $h_+$ and $h_\times$) is decomposed in $(\ell,m)$ multipoles as
\be
\label{eq:hdecomp}
h_+ - \i h_\times = D_L^{-1}\sum_{\ell=2}^\infty\sum_{m=-\ell}^{\ell} h_{\ell m}(t)\,{}_{-2}Y_{\ell m}(\iota,\varphi),
\ee
where $D_L$ is the luminosity distance,
${}_{-2}Y_{\ell m}$ are the $s=-2$
spin-weighted spherical harmonics
and $\iota,\varphi$ are respectively the polar and azimuthal
angles that define the orientation of the binary with respect to the
observer.
Each mode $h_{\ell m}(t)$ can be decomposed
in amplitude $A_{\ell m}(t)$ and phase $\phi_{\ell m}(t)$ as 
\be
\label{eq:hlm}
h_{\ell m}(t) = A_{\ell m}(t)\,\e^{- \i \phi_{\ell m}(t)} \,,
\ee
with a related GW frequency
\be
\label{eq:fgw}
\omega_{\ell m}(t) = \frac{\d}{\d t}{\phi_{\ell m}}(t)\,.
\ee
A dimensionless frequency $\hat{\omega}=GM\omega$
relates to the frequency in Hz according to the formula
\be\non
f = \frac{\omega}{2\pi}\simeq 32.3125\,\hat{\omega}\,\left(\frac{M_\odot}{M}\right)\,\mbox{kHz}\,.
\ee
The GW strain modes are related to the Weyl $\Psi_4$ curvature modes $\psi_\lm$ by
\be
\ddot{h}_\lm = \psi_\lm\,.
\ee
\core~simulations compute only $\psi_\lm$ at different extraction radii $R$. However, the above equation can be integrated to obtain the strain, either by using the fix-frequency integration method \cite{Reisswig:2010di} or directly in the time-domain and performing a polynomial correction, e.g. \cite{Damour:2008te,Baiotti:2008nf,Baiotti:2011am}.
Comparisons between analytical and NR data often use the Regge-Wheeler-Zerilli normalized multipolar waveforms $\Psi_{\ell m}$,
\be
h_{\ell m} = \sqrt{(\ell +2)(\ell +1)\ell (\ell -1)}\,\Psi_{\ell m}\,
\ee
The radiated energy is obtained as~\cite{Bernuzzi:2013rza}
\be
\mathcal{E}_{\rm{rad}} = \frac{1}{16\pi} \sum_{\ell=2}^{\ell_{\rm{max}}}\sum_{m=-\ell}^{\ell} \int_0^t \d t'|D_L \dot{h}_{\ell m}(t')|^2\,,
\ee
whereas the angular momentum is computed as
\be
\mathcal{J}_{\rm{rad}} = \frac{1}{16\pi} \sum_{\ell=2}^{\ell_{\rm{max}}}\sum_{m=-\ell}^{\ell} \int_0^t \d t'm\left[D_L^2 h_{\ell m}(t')\dot{h}^*_{\ell m}(t')\right]\,.
\ee
The data released are computed with $\ell_{\rm{max}}=4$.  The binary dynamics can be characterized by the binding energy and the orbital angular momentum, we therefore work with the binding energy per reduced mass, obtained by substracting the GW energy loss from the initial ADM mass, $E_b = [ (M_{\rm{ADM}}(t=0)- \mathcal{E}_{\rm{rad}}) / M - 1 ]\nu^{-1}$ and the dimensionless rescaled angular momentum $j=(\mathcal{J}_{\rm{ADM}}(t=0)-\mathcal{J}_{\rm{rad}})(M^2\nu)^{-1}$, see \cite{Damour:2011fu,Bernuzzi:2013rza} for details.
The GW luminosity peak is computed as
\be
\label{eq:lpeak}
L_{\rm{peak}}=\max_{t}\frac{\d \mathcal{E}_{\rm{rad}}(t)}{\d t} \approx\max_{t}\left[\frac{1}{16\pi} \sum_{\ell=2}^{\ell_{\rm{max}}}\sum_{m=-\ell}^{\ell} \left| D_L \dot{h}_{\ell m}(t) \right|^2 \right].
\ee
The \textit{moment of merger} is defined as the time of the peak of $A_{22}(t)$, and referred simply as ``merger'' when it cannot be confused with the coalescence/merger process.
Waveforms are often shown in terms of the retarded time
\be
u = t-r_*(r) = t - \left[ r + r_S\ln\left(\frac{r}{r_S}-1\right)\right]\,,
\ee
where $r$ is the coordinate extraction radius in the simulations (assumed close to the isotropic Schwarzschild radius), $r_*$ is the associated tortoise Schwarzschild coordinate, 
 and $r_S=2M$ is the Schwarzschild radius.

\section{Methods}
\label{sec:methods}

\subsection{Initial data}
\label{sec:methods:id}

Initial data for \core~simulations are generated solving Einstein's constraint equations in the conformal thin sandwich (CTS) formalism~\cite{York:1998hy} assuming a helical Killing vector and imposing hydrodynamical equilibrium for the NS's fluid~\cite{Wilson:1995uh,Wilson:1996ty}. It is assumed that the fluid is either irrotational or in a quasi-equilibrium state with constant rotational velocity, which allows for consistent simulations of NS with spin \cite{Tichy:2011gw,Tichy:2012rp}. In the latter formalism, the rotational part $w^a$ of the fluid's velocity is determined by an angular velocity parameter $\omega^i$ as
\be
w_i = \epsilon_{ijk} \omega^j (x^k - x^k_*)\,,
\ee
where $x^k_*$ are the coordinates of the NS center. Possible definitions of spin for a star in a binary are discussed in Refs.~\cite{Bernuzzi:2013rza,Dietrich:2015pxa,Tacik:2015tja}. The spin parameters given in the \core~database are defined to be those of single NSs in isolation with the same rest mass and the same $\omega^i$ as the BNS components \cite{Bernuzzi:2013rza,Dietrich:2015pxa,Tichy:2019ouu}.
To construct initial data with abritrary eccentricities, we use an extension of the helical symmetry condition that is based on approximate instantaneous first integrals of the Euler equations and a self-consistent iteration of the CTS equations \cite{Moldenhauer:2014yaa}. This method also allows us to create low-eccentricity initial data in quasi-circular orbits using an iterative procedure that combines initial data and evolution codes \cite{Dietrich:2015pxa} (see also \cite{Tichy:2009zr,Pfeiffer:2007yz,Kyutoku:2014yba}).

\core~initial data are calculated using either \texttt{Lorene}~\cite{Gourgoulhon:2000nn,Taniguchi:2001qv,Taniguchi:2001ww} or \texttt{SGRID}~\cite{Tichy:2006qn,Tichy:2009yr,Tichy:2011gw,Tichy:2012rp,Dietrich:2015pxa,Tichy:2019ouu}. Both codes use multi-domain pseudospectral methods to solve the CTS equations and surface-fitting coordinates that minimize spurious stellar oscillations at the beginning of the evolutions and guarantee accurate determination of the initial binary global quantities. \texttt{Lorene} can construct irrotational binaries with either piecewise polytropic or tabulated EOS. In the latter case, they are often obtained as cold, $\beta$-equilibrated slices of finite-temperature, composition dependent EOS. \texttt{SGRID} can generate irrotational or spinning binaries with piecewise polytropic EOS and arbitrary (or reduced) eccentricity. In particular, \texttt{SGRID} can simulate BNS in which the individual stars rotate close to the breakup spin and have masses which are $\sim 98\%$ of the maximum supported NS mass allowed by the EOS~\cite{Tichy:2019ouu}. Evolutions of initial data generated with \SGRID{} and \Lorene{} were compared in Ref.~\cite{Bernuzzi:2013rza}, where we found them to be in good agreement.

Initial data in quasi-circular orbits are characterized by the following global quantities of the 3+1 hypersurfaces: the total baryon mass $M_b$ (a conserved quantity along the evolution); the total binary gravitational mass $M$, i.e., the sum of the two gravitational masses of the bodies in isolation; the initial orbital frequency $\Omega\simeq \omega_{22}/2$ and the corresponding ADM mass $M_{\rm ADM}$ and angular momentum $J_{\rm ADM}$. % of the binary at the orbital frequency $\Omega$.

\subsection{Evolution codes}
\label{sec:methods:ev}

\core~simulations evolve initial data using a free-evolution approach to 3+1 Einstein field equations based on the hyperbolic conformal formulations BSSNOK \cite{Nakamura:1987zz,Shibata:1995we,Baumgarte:1998te} or Z4c \cite{Bruegmann:2003aw,Bernuzzi:2009ex,Hilditch:2012fp}. The latter is used for all of the newly released data (Ref.~\cite{Doulis:2022vkx} also uses BSSNOK). The (1+log)-lapse and gamma-driver shift conditions are used for the gauge sector. The general relativistic hydrodynamics is solved in first-order conservative form~\cite{Font:2007zz}. Wave extraction is typically performed on coordinate spheres at finite radius placed in the wave zone of the computational domain (typically $R\sim500-1000~\Msun$) and calculating the Weyl pseudoscalar $\Psi_4$, see e.g. \cite{Brugmann:2008zz} for details.

Simulations are performed with two independent mesh-based codes: \BAM{} \cite{Brugmann:2008zz,Thierfelder:2011yi} and \THC{} \cite{Radice:2012cu,Radice:2013hxh,Radice:2013xpa}, both developed and maintained within our collaboration. These codes employ adaptive mesh refinement (AMR) techniques in which the domain consists of a hierarchy of nested Cartesian grids (refinement levels). The grid spacing of each refinement level in each direction is half the grid spacing of its surrounding coarser refinement level. Finite difference stencils are used for the spatial discretization of the metric variables (usually at fourth order accuracy), and high resolution shock-capturing methods for the hydrodynamics. The Berger-Oliger or Berger-Colella algorithm is employed during the explicit mesh evolution. The latter is performed with the method of lines and Runge-Kutta schemes of third or fourth order accuracy in time. The innermost levels move dynamically during the time evolution following the motion of the NS such that the strong field region around a NS is always covered with the highest resolution. Both codes employ a hybrid OpenMP/MPI parallelization strategy and show good parallel scaling up to thousands of cores.

\BAM{} implements high-order finite-differencing WENO schemes \cite{Bernuzzi:2016pie} and, more recently, an entropy-flux-limited (EFL) scheme~\cite{Doulis:2022vkx}, that is better adapted to the treatment of the NS surface, to accurately simulate multiple orbits and GWs from inspiral-mergers. The typical grid configurations for these simulations consist of seven refinement levels, where the innermost level split into two boxes covering each of the NSs. Standard grid parameters for resolution studies are chosen with $n_m\in[96,256]$ points per direction in each inner (moving) level and $n\in[144,512]$ for the outer levels. The minimal grid spacing in each direction is $\Delta\sim[0.059,0.321]~\Msun$ and the maximal resolution reached in the released simulation is $\Delta\sim 0.059~\Msun$. Symmetries can be imposed to reduce the computational cost of certain problems. For example, aligned-spin BNS are often simulated in bitant symmetry ($z>0$ volume). The simulation parameters can vary for each simulation; the relevant ones are reported in the \core~metadata.

\THC{} implements both high-order finite-differencing schemes \cite{Radice:2015nva} and Kurganov-Tadmor-type central schemes. The latter are preferentially used with simulations with microphysics. \THC{} can make use of microphysical EOS, and implements various neutrino transport schemes \cite{Radice:2016dwd,Radice:2018pdn,Radice:2021jtw} (see below) and subgrid-scale treatment of turbulent mixing and dissipation (GRLES) \cite{Radice:2017zta,Radice:2020ids} to accurately simulate remnants and postmerger dynamics. Most of the GRLES data in the current release employ an effective model for turbulence based on the high-resolution magnetohydrodynamics simulation of Ref.~\cite{Kiuchi:2017zzg}, where the viscosity parameter is set to $\nu_T=\ell_{\rm mix} c_s^2$ and $c_s$ is the sound speed of the fluid. $\ell_{\rm mix}$ is typically defined to be a function of the rest-mass density calibrated with the general-relativistic magneto-hydrodynamics simulations of \cite{Kiuchi:2017zzg} (see \cite{Radice:2020ids}). \THC{} builds on the \Cactus~framework \cite{Goodale:2002a} and the \texttt{Einstein Toolkit} \cite{Loffler:2011ay, EinsteinToolkit}. \THC{} simulations use the \texttt{Carpet} adaptive mesh refinement driver for \Cactus \cite{Schnetter:2003rb}, which implements both vertex centered and cell-centered adaptive mesh refinement with flux correction \cite{Berger:1989a, Reisswig:2012nc}. The grid structure used in the \THC{} simulations is similar to that used in \BAM{}. The grid structure is specified by the resolution at the coarsest refinement level and at the location of the center of the neutron stars. The refinement levels on the grid hierarchy do not have to be connected and \texttt{Carpet} can merge different regions to create grids with complex topology. The standard resolution setup of the \THC{} simulations uses a resolution of $\Delta = 0.125\ M_\odot$ in every direction on the finest refinement level. The maximal resolution reached in the released simulation is $\Delta \simeq 0.08\ M_\odot $. The typical CFL is $0.125$. However, an even lower CFL of $0.0625$ is used on the coarsest grid to handle the gamma-driver source term in the shift evolution equation. All \THC{} simulations included in the current release of the \core~database use bitant symmetry.

\subsection{EOS models}
\label{sec:methods:eos}

\core~simulations currently employ 18 different EOS models for the neutron star matter. \BAM{} data are computed using analytical EOS in the form
\be
P(\rho,\epsilon) = P_{\rm pwp}(\rho) + (\gamma_{\rm th}-1) \rho (\epsilon-\epsilon_{\rm pwp})\,,
\ee
where $P_{\rm pwp}(\rho)$ is a given piecewise politropic EOS model \cite{Read:2008iy}. It prescribes also a value $\epsilon_{\rm pwp}$ for the specific internal energy given the rest mass density $\rho$, augmented with a $\gamma$-law ``thermal'' pressure term (usually, $\gamma_{\rm th}=1.75$~\cite{Hotokezaka:2011dh,Bauswein:2010dn}). The specific parameters we employ for the piecewise polytropic EOS mimic well-established zero-temperature EOS models~\cite{Read:2008iy}; tables of these parameters are available on the \core~website~\footnote{\url{http://www.computational-relativity.org/eos/}}. 

The current release significantly extends the data computed with finite-temperature EOS over the first release. The release includes data from seven finite-temperature EOS, used in the calculation of Refs.~\cite{Radice:2018pdn,Perego:2019adq,Endrizzi:2019trv,Poudel:2020fte,Bernuzzi:2020txg,Camilletti:2022jms,Nedora:2020pak,Prakash:2021wpz}. The finite-temperature EOS include the following models:
BHB$\Lambda\phi$~\cite{Banik:2014qja},
BLh~\cite{Bombaci:2018ksa,Logoteta:2020yxf},
BLQ~\cite{Logoteta:2020yxf,Prakash:2021wpz},
DD2~\cite{Typel:2009sy,Hempel:2009mc},
LS220~\cite{Lattimer:1991nc},
SFHo~\cite{Steiner:2012rk},
SLy4/SRO~\cite{Douchin:2001sv,daSilvaSchneider:2017jpg}.

All these EOS include neutrons, protons, nuclei, electrons, positrons, and photons as relevant thermodynamics degrees of freedom. The ALF2~\cite{Alford:2004pf} and BLQ EOS~\cite{Logoteta:2020yxf,Prakash:2021wpz} are hybrid models accounting for deconfined quark matter. BHB$\Lambda\phi$ is a hadronic model that includes $\Lambda$ hyperons~\cite{Banik:2014qja,Radice:2016rys}.

\begin{figure*}[t]
  \centering
  \includegraphics[width=\textwidth]{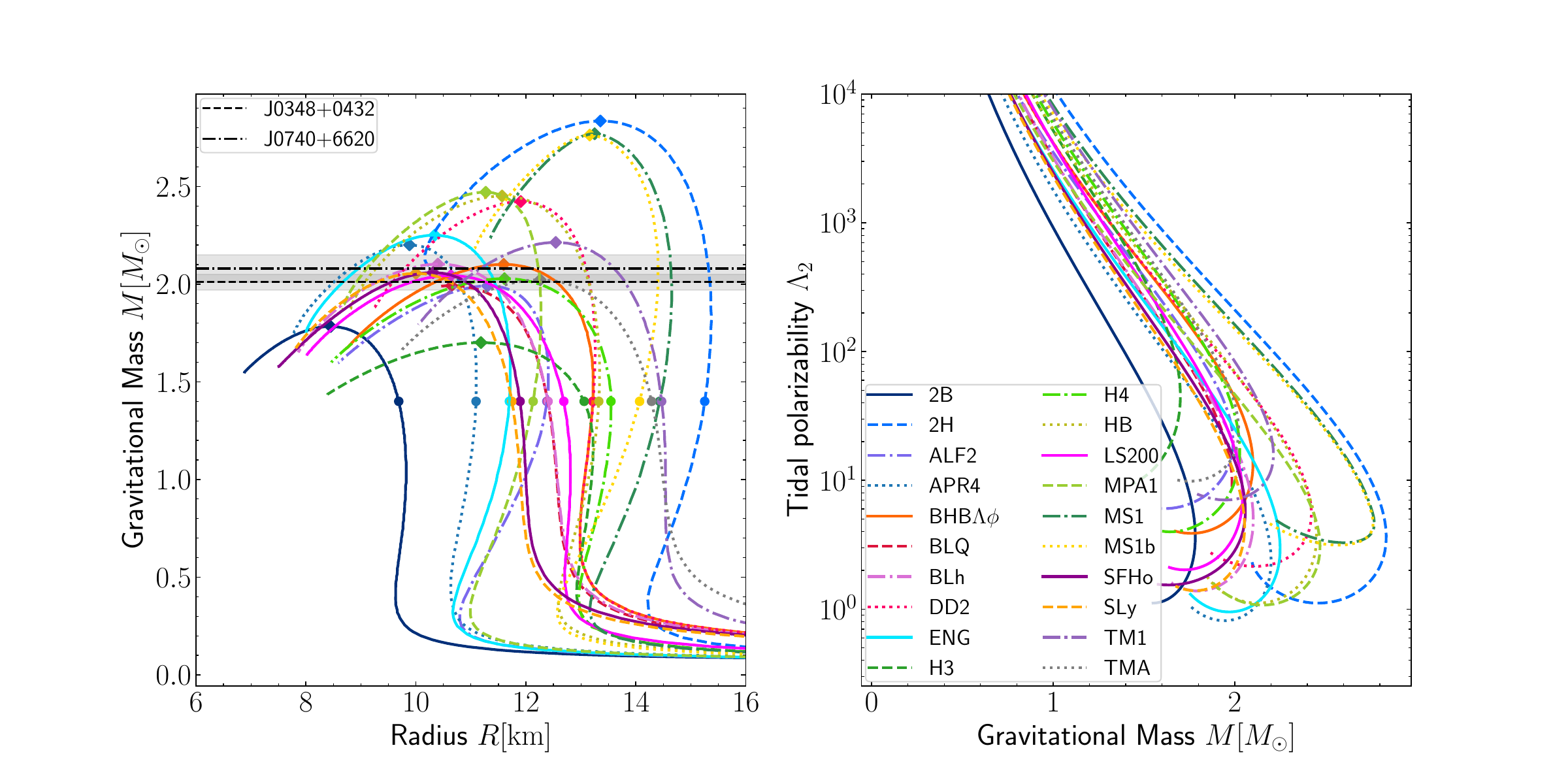}
  \caption{Mass-radius (left) and tidal polarizability-compactness (right) sequences of the available EOS. Diamond-shaped markers correspond to the maximum mass for each EOS, whereas the circle-shaped ones show the radius $R_{1.4}$ for a 1.4~$\Msun$ star. 
  The black lines on the left panel show the current mass estimates of PSR J0348+0432~\cite{Antoniadis:2013pzd} and PSR J0740+6620~\cite{Miller:2021qha,Salmi:2022cgy} with their corresponding constraints (gray bands).}
  \label{fig:MRdiag}
\end{figure*}

Cold, neutrino-less $\beta$-equilibrated matter described by these microphysical EOS predicts NS maximum masses and radii within a larger range than that allowed by current astrophysical constraints, including GW170817 \cite{TheLIGOScientific:2017qsa,Abbott:2018wiz,Abbott:2018exr}. Figure~\ref{fig:MRdiag} shows the mass-radius diagram and the quadrupolar tidal polarizability parameter-mass diagram of these EOS. The largest radius of a $M=1.4~\Msun$ NS is $\R14\sim15.21$~km (EOS 2H) and the smallest radius ${\sim}9.75$~km (EOB 2B). The smallest maximum mass is $\Mmax\sim1.70~\Msun$ (EOS H3), whereas the largest is $\Mmax\sim2.83~\Msun$ (EOS 2H).

Most of these EOS can be found on the \core~website in tabulated form. In the simulations, the EOS is called during the hydrodynamics evolution in order to compute the pressure from the rest-mass density, the temperature, and the electron fraction, i.e., in the form $p=P(\rho,T,Y_e)$.  Any relevant thermodynamical quantity is evaluated by multi-linear interpolating the tabulated values in $\log \rho$, $\log T$, and $Y_e$. As common in relativistic hydrodynamics, the EOS is called during the transformation from conservative to primitive variables. The latter takes place at each timestep and grid point and it involves a numerical root finding of the function $f(p):= p - P(\rho,\epsilon,Y_e)$, where the specific internal energy $\epsilon$ is implicitly given by the temperature $T$ \cite{Galeazzi:2013mia}. Hence, each root-finding step includes another root finder for the function $g(T) = \epsilon-{\cal E}(T)$ (see \cite{Dieselhorst:2021zet} for a discussion on computational efficiency and a non-standard approach based on neural networks.)

\subsection{Microphysics}
\label{sec:methods:micro}

Most of the \THC{} simulations account for the loss of energy and lepton number due to the net emission of neutrinos using a leakage scheme \cite{Galeazzi:2013mia, Radice:2016dwd}. Accordingly, effective neutrino leakage rates are computed as a physically motivated interpolation from the emission and diffusion rates. The latter require the knowledge of the optical depth of each computational zone in such a way as to recover the correct cooling time scale. Neutrino reabsorption is included in some simulations using the M0 scheme \cite{Radice:2016dwd}. This scheme splits neutrinos in an optically thick component, treated with the leakage scheme, and a free-streaming component. The free-streaming neutrinos and their average energies are obtained by solving the radiative transfer equations on a set of radial rays (the so-called ray-by-ray approach) fully-implicitly in time. More recently, we have implemented an energy-integrated M1 scheme in \THC{} \cite{Radice:2021jtw}. The new scheme can self-consistently capture the diffusion of neutrinos from the merger remnant and its reabsorption in the ejecta. M1 simulations are not included in the current release of the database, but will be made public as soon as the associated publications have been accepted. Table \ref{tab:rates} summarizes all neutrino reactions currently included in \THC{} together with the reference in which the form of the rates we use are derived.

\begin{table}
\caption{Weak reaction rates and references for their implementation.
We use the following notation $\nu \in \{\nu_e, \bar{\nu}_e, \nu_{x}\}$
denotes a neutrino, $\nu_{x}$ denote any heavy-lepton neutrino, $N \in
\{n, p\}$ denotes a nucleon, and $A$ denotes a nucleus.}
\label{tab:rates}
\begin{center}
\begin{tabular}{ll}
\hline\hline
Reaction & Reference \\
\hline
$\nu_e + n \leftrightarrow p + e^-$           & \cite{Bruenn:1985en} \\
$\bar{\nu}_{e} + p \leftrightarrow n + e^+$   & \cite{Bruenn:1985en} \\
$e^+ + e^- \rightarrow \nu + \bar{\nu}$       & \cite{Ruffert:1995fs} \\
$\gamma + \gamma \rightarrow \nu + \bar{\nu}$ & \cite{Ruffert:1995fs} \\
$N + N \rightarrow \nu + \bar{\nu} + N  + N$  & \cite{Burrows:2004vq} \\
$\nu + N \rightarrow \nu + N$                 & \cite{Ruffert:1995fs} \\
$\nu + A \rightarrow \nu + A$                 & \cite{Shapiro:1983du} \\
\hline\hline
\end{tabular}
\end{center}
\end{table}

\section{Overview}
\label{sec:overview}

\begin{figure*}[t]
  \centering
  \includegraphics[width=\textwidth]{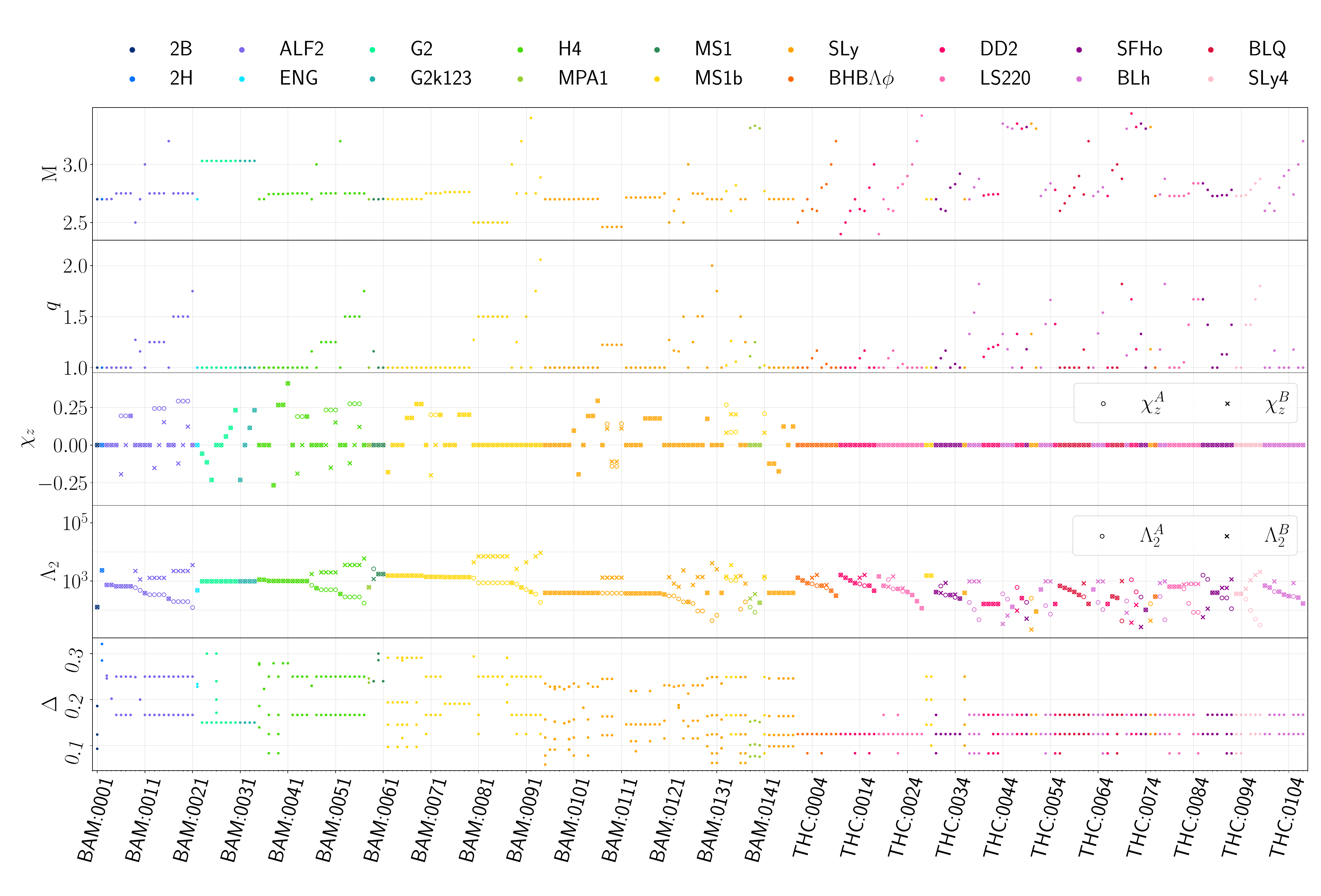}
  \caption{Summary of the \core~database. The complete data are shown
    based on: the binary mass $M$, the mass ratio $q$, the individual
    dimensionless spins $\chi^{\rm{A,B}}_z$, the individual
    quadrupolar tidal parameters $\Lambda^{\rm{A,B}}_2$, the EOS and
    the employed resolution $\Delta$.}
  \label{fig:stats}
\end{figure*}

\begin{figure*}[t]
  \centering
  \includegraphics[width=\textwidth]{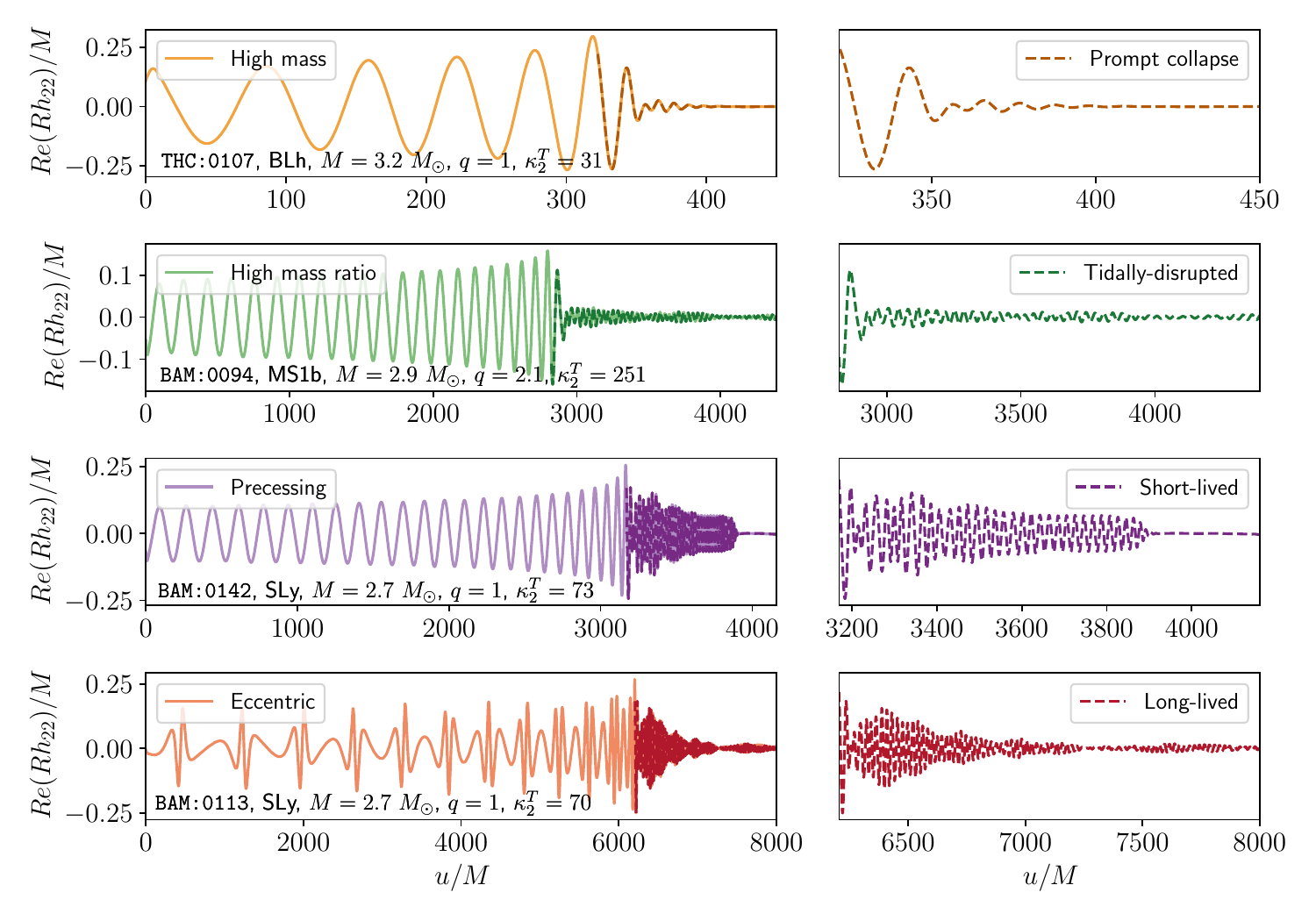}
  \caption{Representative waveforms from the \core~database with their
    respective post-merger signals exemplifying the different
    morphologies influenced by the input physics (total mass of the
    binary, mass ratio, spins, and eccentricity).}
  \label{fig:wvfs}
\end{figure*}

\core~simulations are performed for various binary masses, mass
ratios, NS spins and EOS as summarized by Figure~\ref{fig:stats}. They
cover a significant portion of the BNS parameter space and allow to quantitatively
explore the connection between the gravitational-wave morphology and
the binary parameters in some detail. Figure~\ref{fig:wvfs}
illustrates the variety of waveforms contained in the database. In
the following, we give an overview of the database content and outline 
the connections between physics and waveform morphology.

\begin{figure}[t]
  \centering
  \includegraphics[width=\textwidth]{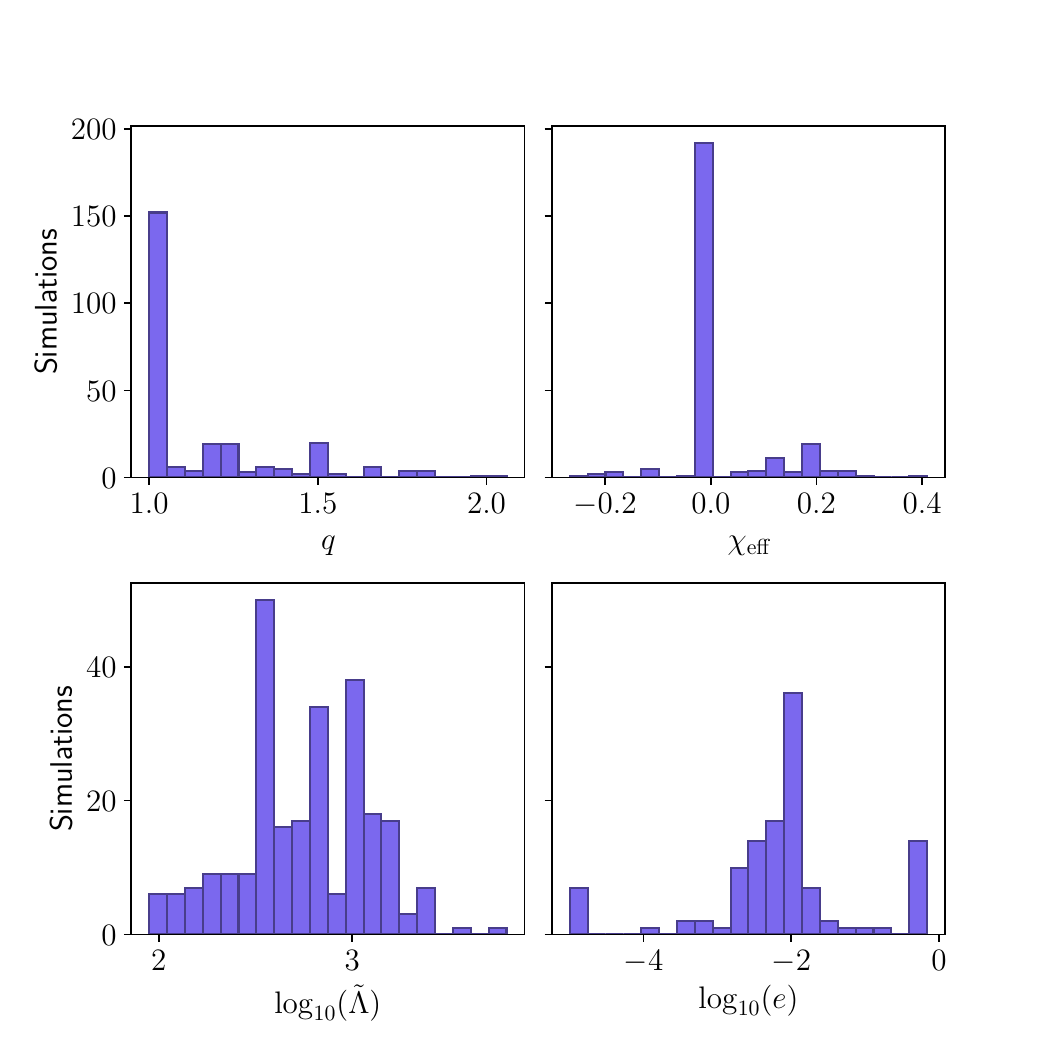}
  \caption{Distribution of the mass ratio $q$, and the effective spin
    parameter $\chi_{\rm eff}$, reduced tidal parameters $\tilde{\Lambda}$,
    and initial eccentricity of the 254 \core~database configurations. Note that for some short simulations no reliable measure of eccentricity was possible.}
  \label{fig:histo}
\end{figure}

The database contains waveforms from binaries with total masses ranging from $2.4~\Msun$ to around ${\sim}3.4~\Msun$ with 45 datasets reaching mass ratios larger than $q\gtrsim1.4$ and up to $q=2.1$ \cite{Dietrich:2016hky,Bernuzzi:2020txg,Ujevic:2022qle}.
EOS effect can be summarized to some extent~\footnote{
A NS spacetime is characterized by an infinite number of multipolar
Love numbers of gravitoeletric and gravimagnetic type; $\Lambda_2$
parametrizes only the (gravitoelectric) leading order term in the
Lagrangian.} by the quadrupolar tidal polarizability parameters
$\Lambda_{1,2}$~\cite{Damour:2009wj}, where larger (smaller) values of
$\Lambda_i$ are associated to stiffer (softer) EOSs. 
The most compact NSs (and most massive binaries) are associated with the smallest
values of $\Lambda_{1,2}$ (and $\tilde{\Lambda}$), see the right panel of
Fig.~\ref{fig:MRdiag}. The \core~data encompasses well the mass and EOS
variation for realistic BNSs. Waveforms from both irrotational and
spinning (using the formalism outlined in Sec.~\ref{sec:methods:id})
quasi-circular mergers are included
\cite{Bernuzzi:2013rza,Dietrich:2016lyp,Chaurasia:2020ntk}. For aligned spins, the individual dimensionless components
range in $\chi_z\in[-0.25,0.5)$; about 7 datasets are from
  simulations with precession effects
  \cite{Dietrich:2016lyp,Chaurasia:2020ntk}. The distribution
  of key parameters among the \core~simulations is shown in
  Fig.~\ref{fig:histo}.

Most of the \core~waveforms are produced from quasi-circular
mergers. The residual eccentricities of non-iterated quasi-circular
initial data is usually $e\sim10^{-2}-10^{-1}$, see the bottom right
panel of Fig.~\ref{fig:histo}. About 13 datasets have an initial
eccentricity $e\lesssim10^{-3}$ that was reduced through an iterative
procedure employing the formalism described in
Sec.~\ref{sec:methods:id}. A subset of waveforms refer instead to
eccentric mergers with initial eccentricity values as high as ${\sim}
0.7$ \cite{Gold:2011df,Radice:2016dwd,Chaurasia:2018zhg}. In
particular, the simulation in the bottom panels of Fig.~\ref{fig:wvfs}
has an initial eccentricity of $0.55$.

\begin{figure}[t]
  \centering
  \includegraphics[width=\textwidth]{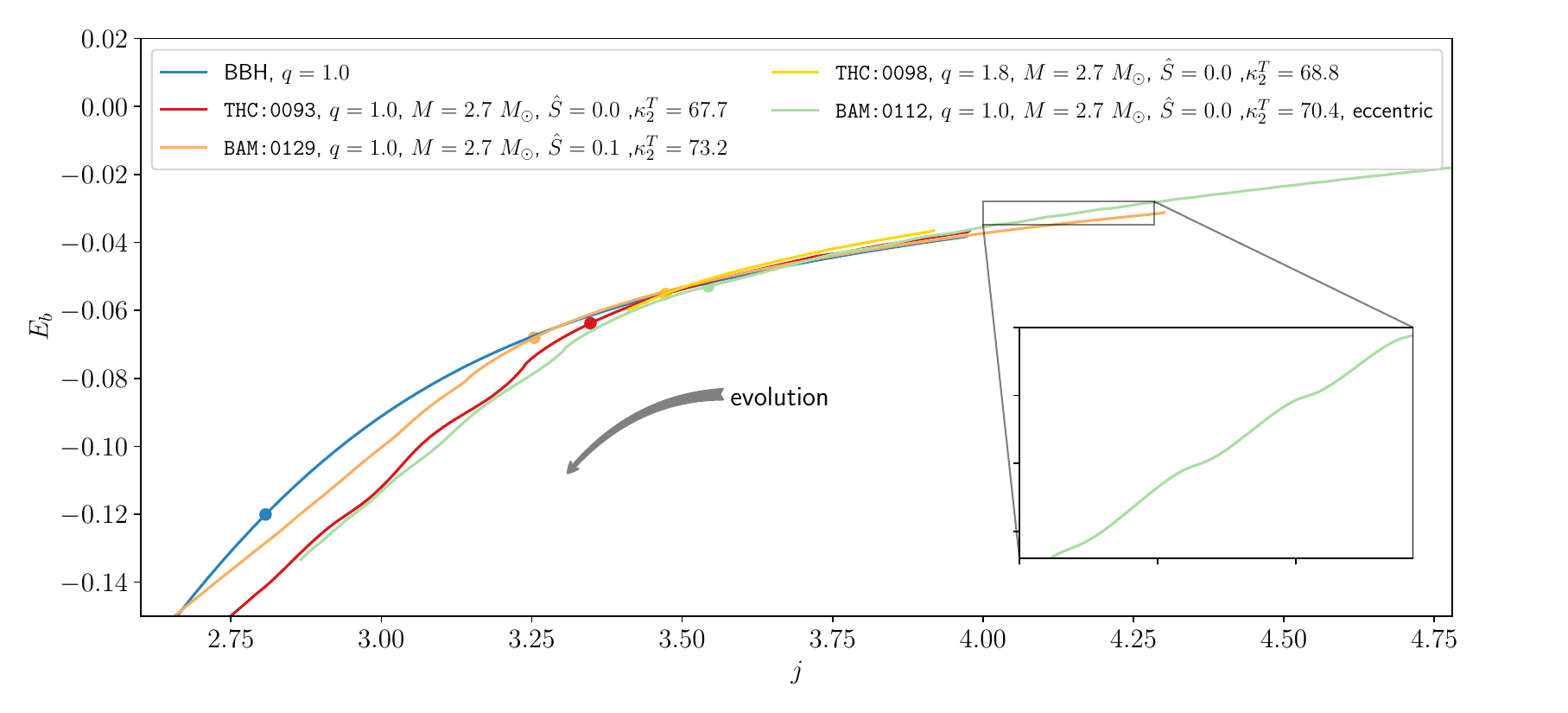}
  \caption{Energy curves $E_b(j)$ for selected binaries.
    The binary's binding energy and angular momentum evolve in time
    from right to left along the cuves $E_b(j)$. The moment of mergers are
    indicated with a marker. The close up of \texttt{BAM:0112} shows a modulation present due to its eccentric inspiral (see text).}
  \label{fig:ej}
\end{figure}

The effects of mass-ratio, spin, and tides on the orbital dynamics can
be studied by means of gauge-invariant energy curves $E_b(j)$, that are also publicly released. 
We illustrate this in Fig.~\ref{fig:ej} for a few examples.
In the inspiral, the binary's angular momentum $j$ decreases due to GW
emission and the system becomes more bound ($E_b$ stays negative and $|E_b|$ increases). Equal
mass ($\nu=1/4$) non-spinning BBH systems merge with $E_b\simeq-0.12$, 
indicating that about 3\% of the binary mass was radiated in GWs to
the moment of merger (marker in the figure).
Tidal effects in BNS make the potential governing the relative
dynamics more attractive. The tidal constribution to the potential at
leading order is $\sim-\kt/r^6$, i.e. it is stronger for larger tidal polarizabilities
$\Lambda_{1,2}$ and it is short-ranged thus affecting the motion
mostly at high frequencies (small separations, $r$) towards merger. Consequently, the inspiral
of an equal mass non-spinning BNS is faster than a binary black hole
inspiral. The binding energy at the moment of merger is
$|E_b|\sim 0.064$, which is smaller than the black hole case because
the BNS system is less compact. Mass-ratio effects make the potential
more repulsive, but are less effective than tides at high
frequencies. The $q=2$ BNS shown in Fig.~\ref{fig:ej} merges at smaller
values $|E_b|\sim 0.055$ than the equal mass because of tidal
disruption. The remnant has also larger angular momentum
$j\sim3.6$~\cite{Nedora:2020pak}.

Spin-effects are dominated by the leading-order spin-orbit interaction;
their character is thus repulsive or attractive depending on the
projection of the spin on the orbital angular momentum
\cite{Damour:2001tu}. This is analogous to what happens to
corotating/counter-rotating circular orbits in Kerr spacetimes that 
move outwards (inwards) for antialigned (aligned) spin
configurations with respect to the nonspinning case.
In binary black hole simulations this effect has been named as
``hang-up'' effect \cite{Campanelli:2006fg}. 
In Fig.~\ref{fig:ej}, the spinning BNS with $\hat{S}=0.1$ is more bound than the
non-spinning BNS at the moment of merger with $E_b\sim-0.068$.
Note that $j$ in this case includes the spin contribution. Moreover, the eccentric equal-mass
case, contrary to the previous ones, shows brief moments of constant $E_b$ indicating the times
when the NSs are apart (see inset of Fig.~\ref{fig:ej}).
Energy curves for BNS have been studied in detail in~\cite{Bernuzzi:2013rza,Dietrich:2016lyp,Chaurasia:2020ntk} to
which we refer for more details.
We stress that the properties of BNS systems at the moment of merger can be
 captured by EOS-insensitive (quasi-universal) relations
\cite{Bernuzzi:2014owa}.
The latter can be helpful in waveform modelling and used to
estimate the properties of the remnant. We refer to Sec.~\ref{sec:QUR} for further discussion.

High-mass BNS produce a remnant that promptly collapses to a black hole shortly after the moment of merger and before the two cores can bounce~\cite{Radice:2020ddv,Bernuzzi:2020tgt}. Prompt collapse implies negligible shocked dynamical ejecta, because the bulk of the mass ejection comes from the first core bounce after their collision \cite{Radice:2018pdn}. Prompt collapse can be characterized by a threshold mass $m_{\rm thr} = k_{\rm thr} \Mmax$, that mainly depends on the maximum mass of cold equilibria $\Mmax$ supported by the EOS \cite{Hotokezaka:2011dh,Bauswein:2013jpa}. The recent analysis of Ref.~\cite{Kashyap:2021wzs}, based on 227 finite-temperature EOS and \core~data~\footnote{ %
These data are not released in the database since the waveforms are rather short and extracted at close radii.
}, found that the prompt collapse mass threshold for equal-mass non-spinning BNS is well described by an EOS-insensitive threshold
\be
k_{\rm thr} = a\,\Cmax + b\,,
\ee
where $\Cmax$ is the compactness of the maximum NS mass, and $a=-3.36\pm 0.20$, $b=2.35\pm 0.06$. A prompt collapse waveform has a rapidly damped black hole ringdown after the moment of merger as shown in the top panels of Figure~\ref{fig:wvfs}. Consequently, the postmerger GW signal is practically negligible for the sensitivities of both current and next-generation detectors. The lack of shocked ejecta and of a massive disc also implies that equal-mass prompt-collapse mergers have dim EM emission. However, for very asymmetric BNS with $q\gtrsim1.4$, it is the tidal disruption of the secondary NS and its accretion onto the primary to trigger the gravitational collapse \cite{Bernuzzi:2020txg}. Thus, asymmetric mergers can be electromagnetically bright because they produce massive tidal dynamical ejecta and remnants with accretion discs of mass ${\sim}0.1~\Msun$. This prompt collapse process is mainly controlled by the incompressibility parameter of nuclear matter around the TOV maximum density \cite{Perego:2021mkd}. A robust, EOS-insensitive criterion is not known in these conditions \cite{Bernuzzi:2020txg,Bauswein:2020aag,Tootle:2021umi,Perego:2021mkd,Kolsch:2021lub}, but tidal disruption effects are subdominant to the mass effect; they produce maximal variations from the equal-mass criterion of ${\sim}8\%$ \cite{Kashyap:2021wzs,Perego:2021mkd}. 

Without prompt collapse, the evolution of a NS remnant is driven by the GWs emission of ${\sim}10^{53}~{\rm erg}$ lasting ${\lesssim}20$~milliseconds (GW-driven phase) \cite{Bernuzzi:2015opx,Zappa:2017xba}. During this phase, a remnant that collapses to a black hole is called {\it short-lived}, while a remnant that settles to an approximately axisymmetric rotating NS  is called {\it long-lived}. Examples of postmerger signals from these remnants are shown in the last two panels on the right of Figure~\ref{fig:wvfs}, for the equal-mass case. The GW-driven phase is associated to a luminous GW transient that peaks at frequencies ${\sim}2-4\,$kHz \cite{Shibata:2002jb,Stergioulas:2011gd,Bauswein:2011tp,Hotokezaka:2013iia,Takami:2014zpa,Bernuzzi:2015rla}. The spectrum of this transient is rather complex but has robust and well-studied features at a few characteristic frequencies. Most of the GW power is emitted in the $(2,2)$ mode at a nearly constant frequency $\omega_{22}(t)\approx 2\pi f_2$; the more compact and close to collapse the remnant is, the higher and more varying the $\omega_{22}(t)$ emission frequency is. The postmerger dynamics is primarily controlled by the masses of the two stars and the bulk properties of the zero-temperature EOS, in particular maximum TOV mass and compactness \cite{Radice:2020ddv,Breschi:2021xrx}. Finite temperature and neutrinos do not produce qualitative differences, other than possibly on the time of gravitational collapse of the remnant~\cite{Zappa:2022rpd}. 
Quantitative differences in the GW signal introduced by finite-temperature and neutrino effects are typically subdominant compared to finite-resolution uncertainties \cite{Sekiguchi:2011zd, Radice:2021jtw}. On the other hand, microphysics plays a crucial role in the EM counterparts and nucleosynthesis from mergers, e.g.,~\cite{Wanajo:2014wha, Sekiguchi:2015dma, Radice:2016dwd, Foucart:2016rxm, Nedora:2020qtd}.

The remnant's signal from asymmetric binaries with mass ratio $q\gtrsim1.4$ carries the imprint of the tidal disruption during merger~\cite{Dietrich:2016hky,Lehner:2016lxy,Bernuzzi:2020txg}.
An example of such a waveform is shown in the second panels (top to bottom) of Figure~\ref{fig:wvfs}. Comparing to the equal-mass long-lived case, the postmerger amplitude is significantly smaller because the asymmetric remnant does not experience the violent bounces of the symmetric remnant. For the same reason, the early-times modulations in frequency and amplitude present in the equal-mass case are significantly suppressed in the asymmetric case.

The evolution of a NS remnant beyond the GW-driven phase is uncertain at present. Explorations of the viscous phase using NR simulations have started \cite{Fujibayashi:2017puw,Nedora:2020pak,Fujibayashi:2020qda}, but they are still incomplete in many ways. While GW emission is expected to be significantly weaker than during merger, remnant's instabilities might enhance GW emission. Current NR results suggest that BNS remnants have an excess of both gravitational mass and angular momentum after the GW-driven phase and when compared to equilibrium configuration with the corresponding baryon mass \cite{Radice:2018xqa,Ciolfi:2017uak}. Possible mechanisms to shed (part of) this energy are CFS \cite{Chandrasekhar:1970b,Friedman:1978b} and one-arm instabilities \cite{East:2015vix,Radice:2016gym,East:2016zvv} that would lead to potentially detectable, long GW transients at ${\lesssim}1\,$kHz. Example of such waveforms are \texttt{THC:0028}, \texttt{THC:0029}, and \texttt{THC:0036} \cite{Radice:2016gym}. 

Finally, \core~data are available for multiple grid resolutions as
discussed in Sec.~\ref{sec:methods} and shown by Fig.~\ref{fig:stats}. Most of
the newly released data contain high resolution simulations with a
minimum grid spacing as low as $\Delta\sim 0.06~\Msun$, e.g., the
NS are resolved with a uniform mesh of spacing
${\sim}88.4$~meters. Notably, simulations of more than 20 orbits or
up to hundreds milliseconds postmerger and with microphysics were
performed at these resolutions. Simulations at multiple resolutions are a crucial aspect for data quality that is discussed next.

\section{Waveform accuracy}
\label{sec:acc}

Waveform accuracy depends on several aspects of the simulations. Within the \core~data the largest sources of uncertainty are (i) the truncation error of the numerical scheme, that is regulated by the mesh resolution employed in the simulations, and (ii) the finite extraction radius for the GW data, e.g.~\cite{Bernuzzi:2011aq,Radice:2013hxh,Bernuzzi:2016pie,Doulis:2022vkx}. Other aspects are relevant for waveform modelling, as for example, the length of the simulation (number of orbits/GW cycles), the residual eccentricity in quasi-circular initial data, and the simulation of realistic physics (star rotation, EOS, etc.).

Waveform accuracy should be studied by the user case-by-case considering amplitude and phase plots with datasets of simulations at different resolutions and extraction radii. This analysis typically requires a minimum of three simulations of the same BNS at different grid resolutions (a ``convergent series'') and has been performed by the authors in Refs. \cite{Bernuzzi:2011aq,Radice:2013hxh,Radice:2013xpa,Radice:2015nva,Bernuzzi:2016pie,Kiuchi:2017pte,Kiuchi:2019kzt,Doulis:2022vkx}. We give below in Sec.~\ref{sec:acc:phasing} a complete example of error analysis of a ${\sim}10$ orbit inspiral-merger waveform.

In GW astronomy, the quality of a waveform template is commonly assessed using the Wiener product (\textit{overlap}) between two waveforms $h_{1,2}(t)$ for a given detector~\cite{Cutler:1994ys},
\be\label{eq:psd}
\langle h_1,h_2\rangle := 4\Re{ \int \frac{\tilde{h}_1(f)\tilde{h}^*_2(f)}{S_n(f)} }df\,,
\ee
where $S_n(f)$ is the power spectral density (PSD) 
of the detector noise and $\tilde{h}_{1,2}(f)$ the Fourier transform of $h_{1,2}(t)$. The inner product allows to define ``accuracy standards'' for either detectability or measurements (parameter estimation), e.g., Ref.~\cite{Apostolatos:1995pj,Owen:1995tm,Damour:1997ub,Lindblom:2008cm,Lindblom:2009ux,Lindblom:2010mh,Damour:2010zb}. In the former case, one is interested in quantifying the fractional loss of signal-to-noise ratio (SNR) due the use of a sub-optimal, discrete match filter. Since the number of GW events is proportional to the observable volume, and the distance is inversely proportional to the observed SNR, the fractional loss of potential events scales like the cube of the minimum overlap in the discrete template bank \cite{Apostolatos:1995pj,Owen:1995tm,Damour:2010zb}. In the latter case, one is interested in quantifying the bias (or the maximum knowledge) on the GW parameters given the noise in the detector (statistical errors) \cite{Damour:1997ub,Lindblom:2008cm,Damour:2010zb}. In practice, one proceeds by defining the \textit{faithfulness} between two waveforms
\be\label{eq:faith}
\mathcal{F}:=\mathop{\rm{max}}_{t_0,\phi_0}\frac{\langle h_1,h_2\rangle}{\|h_1\| \|h_2\|},
\ee
where $t_0$ and $\phi_0$ are a reference initial time and phase, and its complementary, the unfaithfulness, $\bar{\mathcal{F}}:= 1-\mathcal{F}$.
By demanding that, at worst, the systematics biases become of the same order as the statistical ones when the noise level is doubled, it is possible to establish the condition \cite{Damour:2010zb}
\be
\label{eq:fthreshold}
\mathcal{F} > 1 - \frac{\epsilon^2}{2 \rho^2}\,,
\ee
where $\rho$ is the SNR and $\epsilon^2\ll1$. This condition is \textit{necessary} for unbiased parameter estimation (faithful waveforms); its violation does not imply that an analysis has biases~\cite{Flanagan:1997sx,Lindblom:2008cm,Bernuzzi:2011aq,Gamba:2020wgg}. The above criterion can be used to quantify the accuracy of NR data, for example by calculating the faithfulness between data at different resolutions~\cite{Bernuzzi:2011aq,Gamba:2020wgg,Doulis:2022vkx}. We will use the faithfulness measure in Sec.~\ref{sec:acc:FF} to discuss the average accuracy of the data of the \core~database.

\subsection{Example of NR waveform analysis}
\label{sec:acc:phasing}

\begin{figure}[t]
   \centering
     \includegraphics[width=\textwidth]{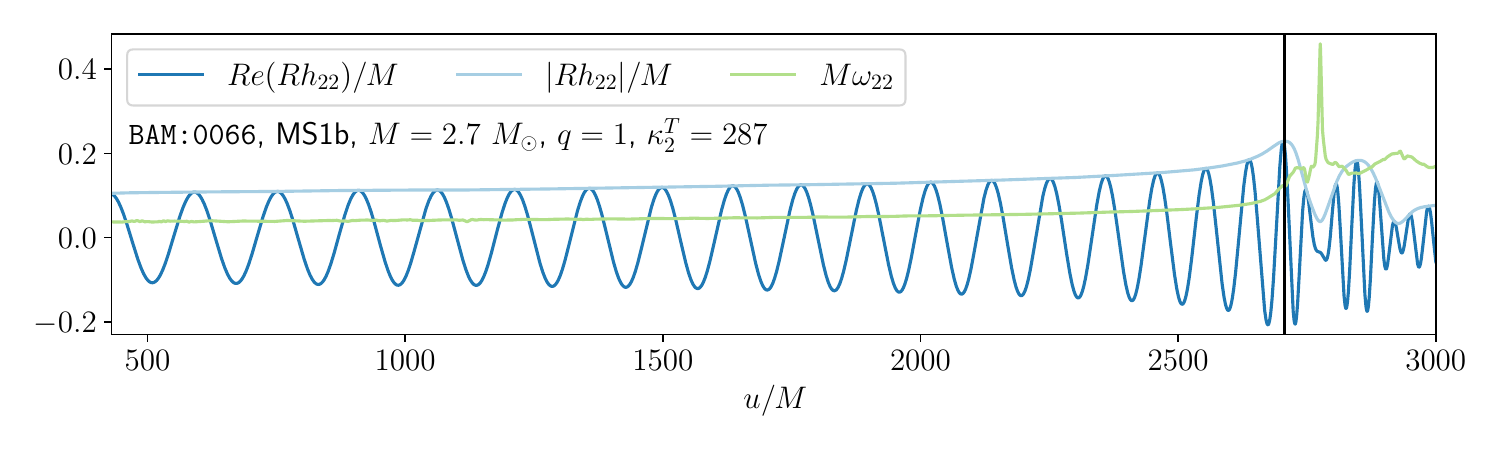}
     \caption{The real part of the strain $Rh_{22}$, its amplitude $|Rh_{22}|$, and frequency $M\omega_{22}$ of \texttt{BAM:0066}. The black solid line indicates the moment of merger.}
      \label{fig:bamwvf}
\end{figure}

In this section we present a waveform error analysis for
\texttt{BAM:0066}~\cite{Dietrich:2017aum}. This example effectively represents data that exhibit second order convergence. Figure~\ref{fig:bamwvf} shows the strain $Rh_{22}$ at the
lowest extraction radius available for this simulation,
$R=700\,\Msun$, its amplitude $|Rh_{22}|$ and frequency $M\omega_{22}$. Note that in this section we
use $R$ instead of $D_L$.

\begin{figure}[t]
   \centering
     \includegraphics[width=\textwidth]{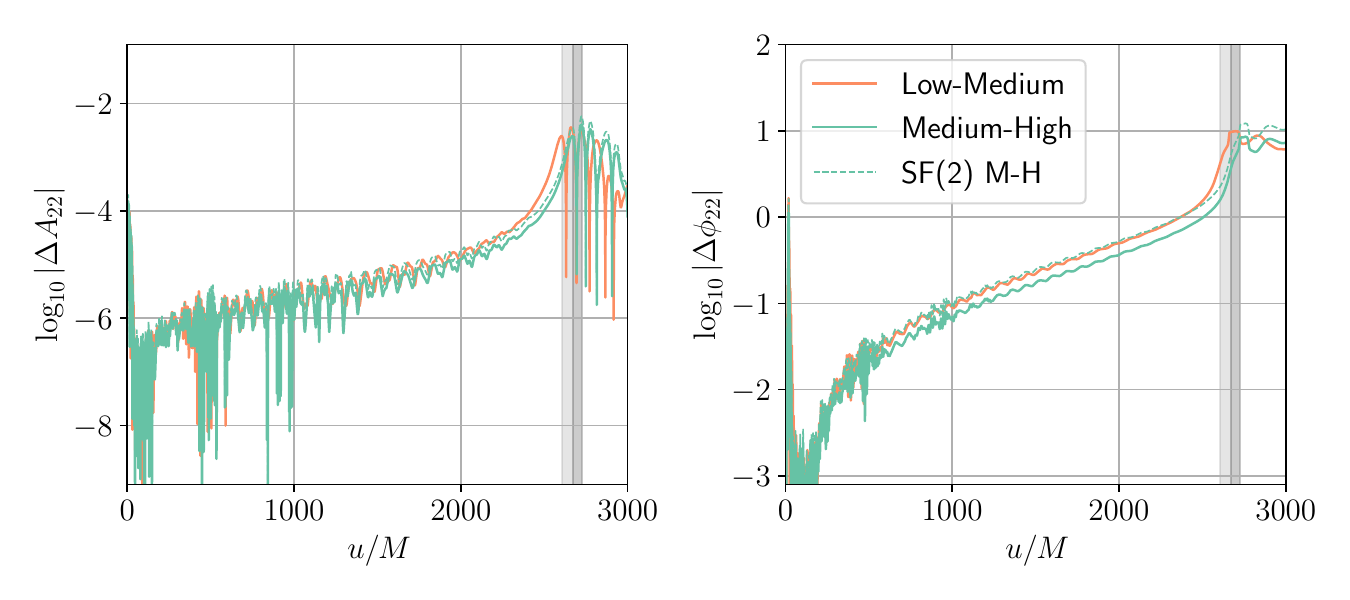}
     \caption{Self-convergence tests of \texttt{BAM:0066}
       $R\psi_{22}$        The dashed blue line represents the rescaled difference for second order convergence. The light (dark) gray shaded regions show the time of merger differences of Low-Medium (Medium-High) resolutions. }
  \label{fig:bamconv}
\end{figure}

In order to test self-convergence, we compare amplitude and phase differences of $R\psi_{22}$ between the different resolutions. For this case, we consider the simulation at resolutions $n_m=120,160,240$ grid points on the highest refined AMR level;  hereafter Low (L), Medium (M), and High (H). The convergence rate $p$ is found experimentally by rescaling these differences using the scaling factor ${\rm SF}$~\cite{Bernuzzi:2011aq},
\be
{\rm SF} =\frac{\Delta^p_{\rm L} - \Delta^p_{\rm M}}{\Delta^p_{\rm M} - \Delta^p_{\rm H}}
\ee
where $\Delta_x$ is the grid spacing at resolution $x$. We show the self-convergence test in Figure~\ref{fig:bamconv}. The differences decrease with increasing resolution, as one would expect from convergent data. They also increase with increasing simulation time because truncation errors accumulate during the simulation. The optimal scaling is found for $p=2$ with ${\rm SF}(2)=1.4$, thus indicating second order convergence. In presence of convergence, a measure of the error to be assigned to the (highest resolution) data is given simply by the difference between the two highest resolutions. This is a conservative estimate because (for convergent data) the truncation error is certainly smaller. Alternatively, the experimental convergence factor can be in principle used in a Richardson extrapolation of the data to provide an improved dataset and error estimate \cite{Bernuzzi:2011aq,Bernuzzi:2016pie}. Note that in this procedure the waveforms are not shifted by a relative time and phase shift because the simulations of the convergent series are run using the same initial data with a fixed initial phase.

\begin{figure}[t]
   \centering
   \includegraphics[width=\textwidth]{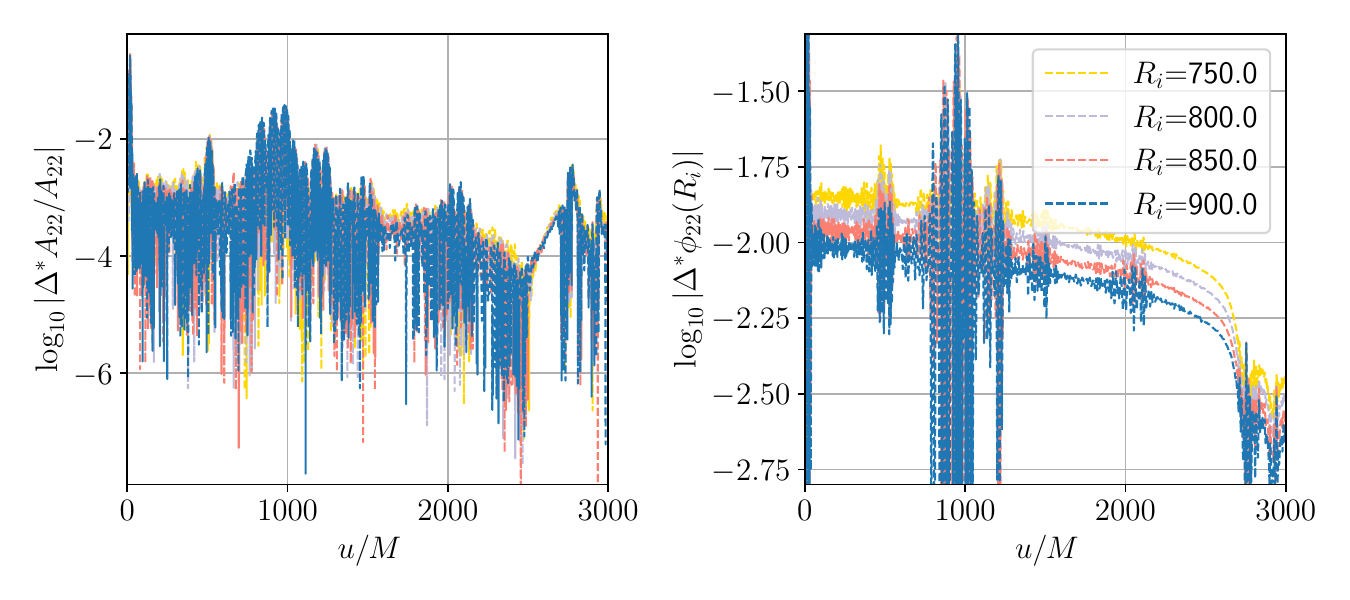}
           \caption{Amplitude (left) and phase (right) differences
         between \texttt{BAM:0066}'s $R\psi_{22}$          extracted at consecutive finite-radii $R_i=700, 750, 800, 850, 900~\Msun$.}
     \label{fig:bamstar}
\end{figure}

To assess the uncertainties originated from the waveform obtained at finite-extraction radii, $R_i$, we compare the phase differences between consecutive radii~\cite{Bernuzzi:2011aq}
\be
\Delta^*\phi_{22}(R_i)=\phi_{22}(R_i)-\phi_{22}(R_{i-1})\,,
\ee
and similarly for the relative amplitudes, $\Delta^*A_{22}/A_{22}$. In Fig.~\ref{fig:bamstar} we show the differences at the extraction radii $R= 700, 750, 800, 850, 900~\Msun$.
The phase differences decrease at progressively large radii, thus indicating the numerical waveforms are converging towards their true morphology at null infinity. The phase differences are larger at early times and decrease towards merger; note this behaviour has the opposite sign of that of resolution effects \cite{Bernuzzi:2016pie}. The relative differences in amplitude are ${\sim}10^{-4}$ for all radii, indicating robust results are obtained already with relatively close extraction sphere.
The waveforms can be extrapolated to null infinity using either a
polynomial in $1/R$ of order $K$ \cite{Bernuzzi:2011aq} or the method outline in
\cite{Lousto:2010qx}. The two methods give comparable results; the
former is more general and can be applied to the curvature multipoles
$\psi_\lm$, the latter is a simpler method for the strain modes. An
error due to finite extraction can be then assigned to the data at
finite extraction as the difference with the extrapolated data (or
viceversa). Another method is to post-process simulations using Cauchy
characteristic extraction (CCE)~\cite{Reisswig:2009rx} and to simulate
the waveform at future null infinity. This technique was used for some of the \core~data.

\begin{figure}[t]
   \centering
     \includegraphics[width=0.8\textwidth]{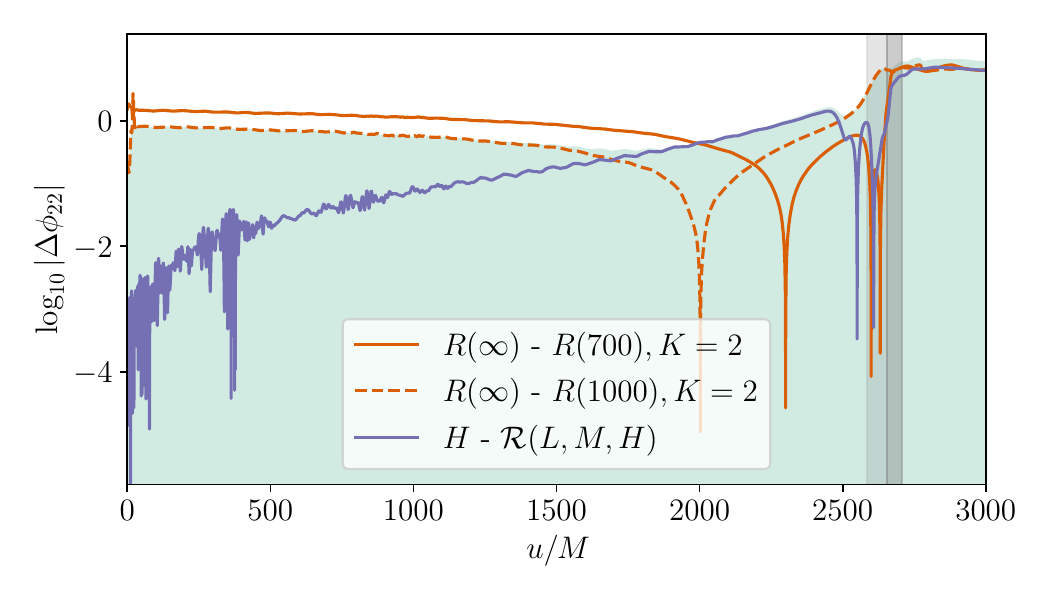}
     \includegraphics[width=0.8\textwidth]{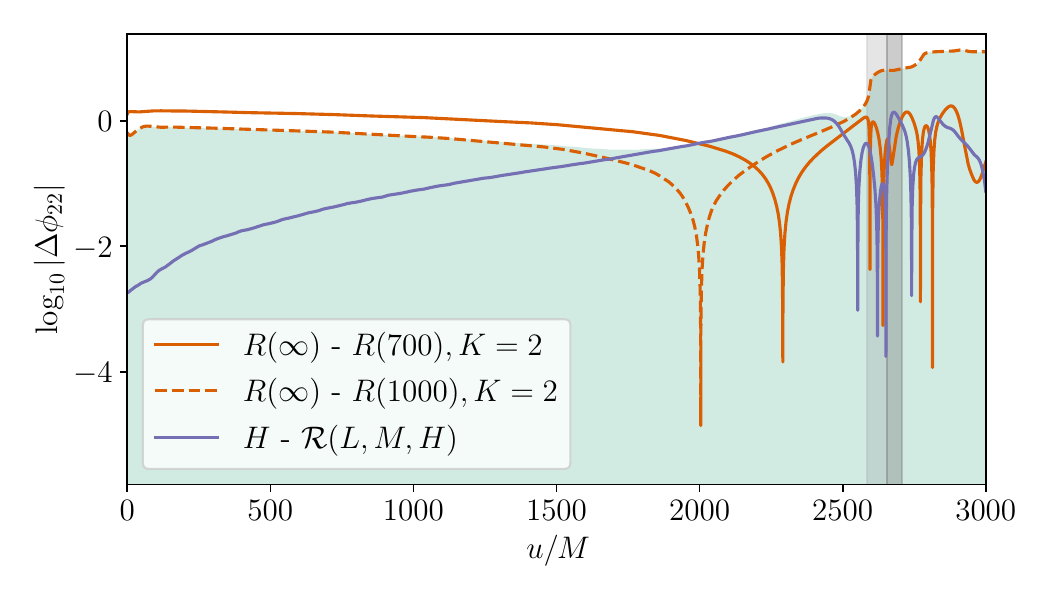}
     \caption{Error budget (shown as a green shaded area) for the phase of $R\psi^4_{22}$ (top) and $Rh_{22}$ (bottom) from the truncation error (blue line) and finite-radius extraction uncertainty (orange lines) employing polynomial extrapolation with $K=2$. The blue line shows the difference between the highest resolution and the Richardson extrapolated dataset $\mathcal{R}(L,M,H)$.}
  \label{fig:errorbud}
\end{figure}

 The total error budget can be computed as the sum in quadrature of the
 truncation and finite extraction errors, and it is shown in
 Fig.~\ref{fig:errorbud} for both the curvature and strain (2,2)
 modes. As mentioned above, the truncation phase error is typically a
 factor ${\sim}2$ larger than the finite extraction error (for
 $R\gtrsim500\Msun$) at merger and in simulations with tens of
 orbits.

\begin{figure}[t]
   \centering
     \includegraphics[width=0.6\textwidth]{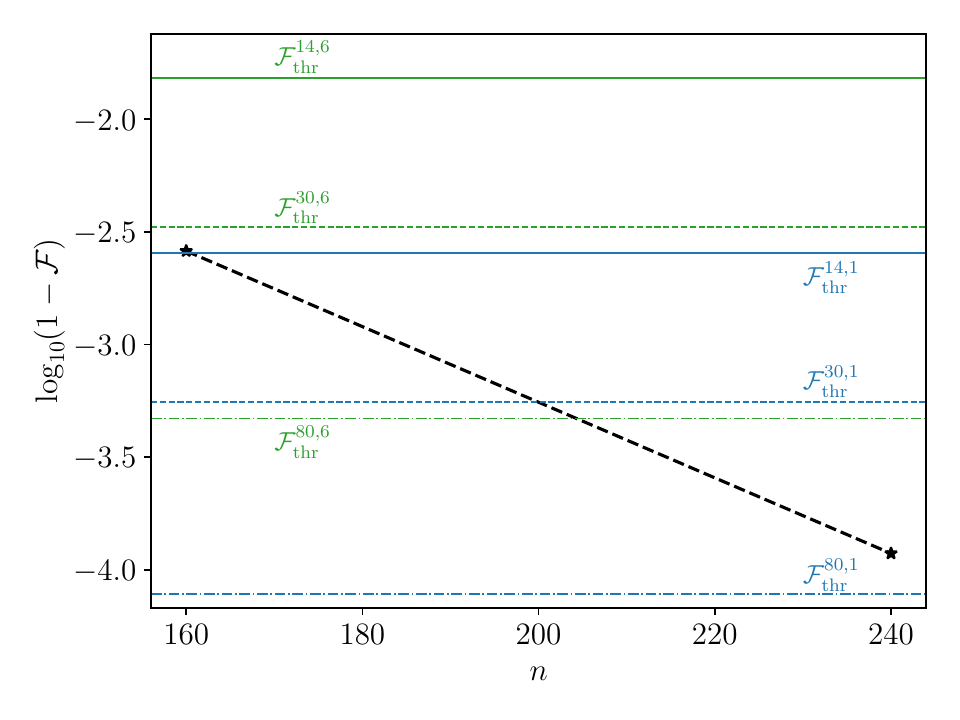}
     \caption{Unfaithfulness of \texttt{BAM:0066}'s Medium-High ($n_m=240,160$) and Low-Medium ($n_m=160,120$) $Rh_{22}$ waveforms. The blue and green lines represent the accuracy tests $\mathcal{F}^{\rho,\epsilon^2}_{\rm thr}$ for different values of $\rho$ (SNR) and $\epsilon^2$.}
  \label{fig:bammat}
\end{figure}

Finally, we obtain the unfaithfulness $\bar{\mathcal{F}}$ of the waveforms between the different resolutions (M-H and L-M). The Wiener integral is evaluated in the frequency range $f\in[f_{\rm min},f_{\rm mrg}]$ and employing the Advanced LIGO PSD P1200087~\cite{aLIGODesign_PSD} from \texttt{bajes}~\cite{Breschi:2021wzr}. 
Here $f_{\rm min}$ corresponds to the initial GW frequency, and $f_{\rm mrg}$ to the frequency at the moment of merger. For the faithfulness threshold $\mathcal{F}_{\rm thr}$ in Eq.~(\ref{eq:fthreshold}), we consider $\epsilon^2=1$ as the strict requirement, and $\epsilon^2=6$, corresponding to the number of intrinsic parameters of a BNS. Similarly to \cite{Doulis:2022vkx}, the SNR values are chosen to be $\rho=14, 30, 80$. Figure~\ref{fig:bammat} shows the computed values. The smallest unfaithfulness (M-H, $n_m=240,160$) passes five out of the six accuracy tests, whereas the other one (L-M, $n_m=160,120$) passes only two, namely $\mathcal{F}^{14,6}_{\rm thr}$ and $\mathcal{F}^{30,6}_{\rm thr}$. However, the unfaithfulness value lies closely (or on top) of the threshold $\mathcal{F}^{14,1}_{\rm thr}$.

Analyses similar to the one above are necessary to determine the quality of the NR data for GW modelling. Convergence of the data is a necessary requisite for robust error estimates. Other diagnostic quantities used to verify convergence in simulations are constraint violation, baryon mass conservation and the stars oscillations during the first orbits, e.g. \cite{Thierfelder:2011yi,Bernuzzi:2011aq,Hilditch:2012fp,Dietrich:2015pxa,Chaurasia:2020ntk}. Achieving waveform convergence in long-term evolutions of BNS is a nontrivial result and, in our experience, requires at least fifth order finite-differencing schemes or finite volume schemes with fifth order reconstructions (at the current resolutions)~\cite{Radice:2013xpa,Bernuzzi:2016pie,Doulis:2022vkx}. Second~\cite{Bernuzzi:2016pie}, approximately third~\cite{Radice:2013xpa} and clear fourth order convergence~\cite{Doulis:2022vkx} has been demonstrated up to merger in some data using these finite-differencing conservative schemes. Extreme mass ratios $q\sim2$ and NS rotation close to the breakup limit remain challenging to simulate as well as to obtain clean convergence in GW higher (subdominant) modes like $(\ell,m)=(2,1), (3,3)$ and $(4,4)$. Work in these directions is ongoing~\cite{Bernuzzi:2020txg,Dudi:2021abi,Ujevic:2022qle,Doulis:2022vkx}. For example, clear fourth order convergence in the subdominant (3, 2) and (4, 4) modes for $q=1$ has been shown in~\cite{Doulis:2022vkx}. Postmerger waveforms typically show slower convergence due to shock formation at merger and the complex fluid dynamics in the remnant. Nonetheless, GW spectra have remarkably robust features that can be accurately quantified with NR data, as we shall discuss in Sec.~\ref{sec:QUR}. We refer the reader to Ref.~\cite{Breschi:2019srl,Breschi:2022xnc} for recent work on the accuracy of \core~postmerger waveform.

\subsection{Faithfulness analysis}
\label{sec:acc:FF}

In an attempt to give an overview of the accuracy of the waveform
database, we compute the unfaithfulness of the (2,2) mode waveforms
$h_{22}$ between the highest and second highest resolutions, for the
whole database. 
We use again the zero-detuned, high-power Advanced LIGO
PSD~\cite{aLIGODesign_PSD}. The minimum frequency $f_{\rm min}$
employed in the integral of Eq.~(\ref{eq:psd}) corresponds to the
initial frequency of each individual simulation. 

\begin{figure}[t]
   \centering
        \includegraphics[width=\textwidth]{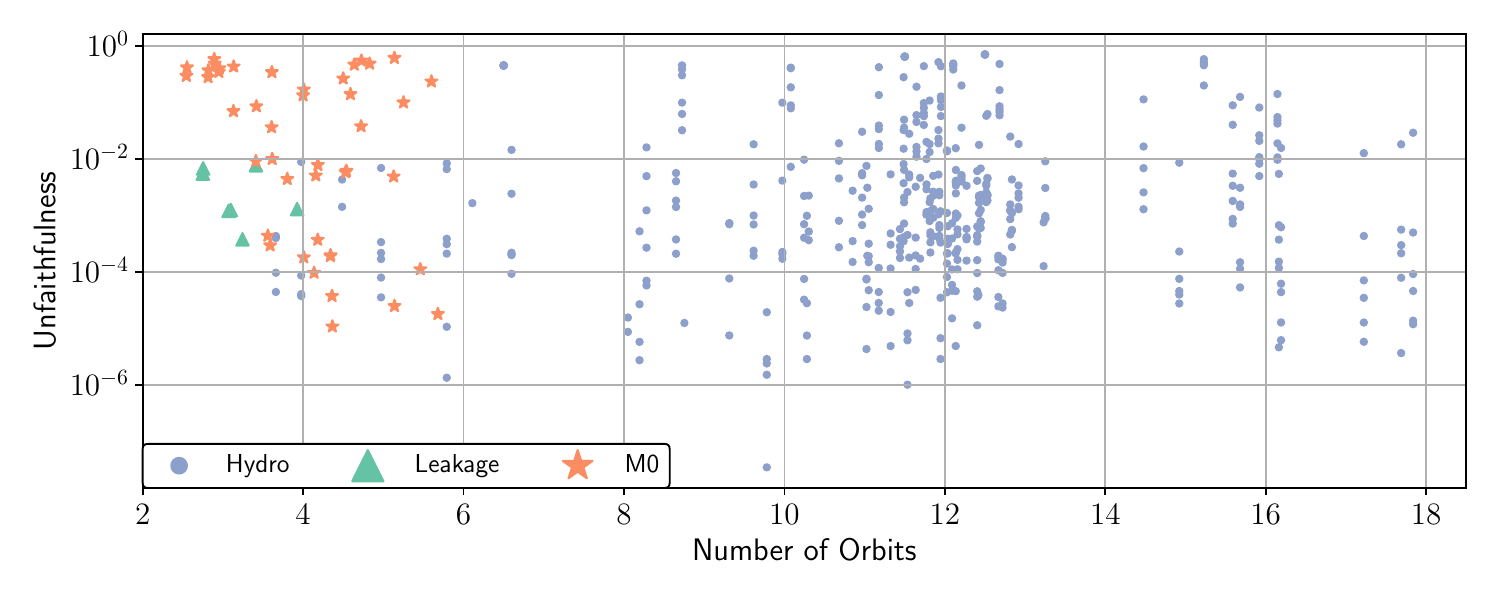}
     \caption{Unfaithfulness computed between the highest and second
       highest simulations $h_{22}$ waveforms for every configuration
       of the database. The different colors and markers correspond to the
       microphysics scheme employed for each simulation.}
     \label{fig:mismatches}
\end{figure}

The result of this analysis is summarized in Fig.~\ref{fig:mismatches},
where $\bar{\mathcal{F}}$ is shown as a
function of the number of orbits and different colors mark the
microphysics scheme employed in each simulation.
The unfaithfulness values are scattered on a wide range, but about 65\% of
the waveforms lay below the 1\% level which is conventionally
considered the accuracy threshold for detection purposes. Importantly, the
dependence on the number of orbits (simulation length) is very weak
and most of the simulations with ten or more orbits have
$\bar{\mathcal{F}}<0.01$. Several waveforms from multiple-orbits have
$\bar{\mathcal{F}}\lesssim10^{-4}$; according to the analysis in the
previous section, these data can be considered faithful (suitable for
parameter estimation) up to signal SNR of 30-80.
We note that data with very few orbits (e.g. \texttt{THC:0019},
\texttt{BAM:0029}, and \texttt{BAM:0082}) show a remarkably low
unfaithfulness. These simulations have a short inspiral and
rather focus on the postmerger signal, which is not considered in this
analysis. Hence, small $\bar{\mathcal{F}}$ is not necessarily an
indication that these simulations are suitable for waveform modelling.

A faithfulness analysis for postmerger signals was recently
presented  by some of us in \cite{Breschi:2019srl,Breschi:2022xnc}. There, we
found average mismatches of ${\sim}0.01-0.4$. The main source of
uncertainty in the postmerger waveforms is the numerical resolution
(see the above Section) and the impact of the 
resolution on the remnant's collapse.

\section{Quasi-universal relations}
\label{sec:QUR}

\begin{table}
 \caption{\label{tab:fits}Updated fit coefficients for relevant merger and PM quantities.}
 %\footnotesize\rm
 \scriptsize\rm
 \begin{tabular*}{\textwidth}{@{}l*{15}{@{\extracolsep{0pt plus12pt}}c}}
 \br
 $Q^{\rm fit}$ & $a_0$ & $k$ &$a^M_k$ &$a^S_k$ &$b^S_k$ &$a^T_k$ &$b^T_k$ & $\chi^2$ & Error & $R^2$\\
 \mr
 \multirow{4}*{$A_{\rm mrg}/M $} & \multirow{4}*{0.55} & 1 & 5.27 & 0.31 & -39.21 & 5.59$\times 10^{-2}$ & -2.51$\times 10^{-2}$ & \multirow{4}*{0.113} & \multirow{4}*{2.6\%}& \multirow{4}*{0.949}\\
 & & 2 & & && 1.00$\times 10^{-6}$& -2.00 \\
 & & 3 & & && 0.12 & 11.09  \\
 & & 4 & & && 6.79$\times 10^{-5}$ & 9.72  \\
 \hline
 \multirow{4}*{$Mf_{\rm mrg}/\nu$} & \multirow{4}*{0.22} &1 & 0.80 & 0.25 & -1.99 & 4.85$\times 10^{-2}$ & 1.80 & \multirow{4}*{0.329} & \multirow{4}*{4.5\%} & \multirow{4}*{0.925}\\
 & & 2 & & &  & 5.86$\times 10^{-6}$ & 599.99\\
 & & 3 & & &  & 0.1 & 7.80\\
 & & 4 & & &   & 1.86$\times 10^{-4}$ & 84.76  \\
 \hline
 \multirow{4}*{$Mf_2$} & \multirow{4}*{8.99$\times 10^{-2}$} & 1 & 31.02 & 7.42$\times 10^{-2}$ & 29.99 & 2.94$\times 10^{-2}$ & 1.13 & \multirow{4}*{0.067} & \multirow{4}*{3.6\%} & \multirow{4}*{0.958}\\
 & & 2 & & & & 3.78$\times 10^{-5}$  & -0.99\\
 & & 3 & & & & 5.75$\times 10^{-2}$ & 39.99 \\
 & & 4 & & & & 2.77$\times 10^{-4}$  & 27.77 \\
 \br
 \end{tabular*}

 \caption{\label{tab:fits_Mf2Rx}Updated fit coefficients for $Mf_2$ as a function of the NS radii $R_{1.4}$ and $R_{1.8}$.}
  %\footnotesize\rm
  \scriptsize\rm
  \begin{tabular*}{\textwidth}{@{}l*{15}{@{\extracolsep{0pt plus12pt}}c}}
    \br
 				 & $a_0$ & $a_1$ & $a_2$ & $a_3$ & $\chi^2$ & Error & $R^2$\\
    \mr
$Mf_2(R_{1.4}/M)$ 				& 0.24 & -0.10 & 1.13$\times 10^{-2}$ & - & 0.55 & 5.9$\%$ & 0.901  \\
$Mf_2(R_{1.8}/M)$                & 0.23 & -0.10 & 1.21$\times 10^{-2}$ & - & 0.31 & 4.5$\%$ & 0.949 \\
\mr
$Mf_2(R_{1.4}/M,R_{1.4}/R_{1.8})$& 0.15 & -0.11 & 1.38$\times 10^{-2}$ & 9.76$\times 10^{-2}$ & 0.31 & 4.5$\%$ &  0.949 \\
$Mf_2(R_{1.8}/M,R_{1.4}/R_{1.8})$& 0.20 & -0.10 & 1.22$\times 10^{-2}$ & 2.77$\times 10^{-2}$ & 0.30 & 4.4$\%$ & 0.952  \\
    \br
  \end{tabular*}

  \caption{\label{tab:fits_lp}Best fit coefficients for the luminosity peak. The last columns show the $\chi^2$, the fit's relative standard deviation and the coefficient of determination $R^2$.}
  %\footnotesize\rm
  \scriptsize\rm
  \begin{tabular*}{\textwidth}{@{}l*{15}{@{\extracolsep{0pt plus12pt}}c}}
    \br
 				 & $k$ & $p_{k10}$ & $p_{k11}$ & $p_{k20}$ & $p_{k21}$ & $p_{k30}$ & $p_{k31}$& $\chi^2$ & Error & $R^2$\\
    \mr
 				& 1 & 2.28 & 7.59$\times 10^{-1}$ & -17.74 & -0.57 & -17.47 & 4.58  \\
$L_{\rm peak}/\nu$ & 2 & -8.38$\times 10^{-2}$ & 9.61$\times 10^{-3}$ & 3.24$\times 10^{-1}$ & -3.33$\times 10^{-2}$ & 13.91 & 10.10   & 2.23 & 12\% & 0.961\\
  				& 3 & -5.18$\times 10^{-1}$ & 14.64 & -5.35 & -50.54 & 11.61 & -29.96  \\
    \br
  \end{tabular*}
\end{table}

As a first application of the database, we present in this section new EOS-insensitive relations for the merger and postmerger waveforms. Previous work found that several key quantities characterizing the merger dynamics depend on the unknown EOS mainly throughout the tidal parameters and have a very weak dependence on other details of the matter model, e.g.,~\cite{Bauswein:2011tp,Read:2013zra,Bernuzzi:2014kca,Bauswein:2015yca,Rezzolla:2016nxn,Zappa:2017xba,Bernuzzi:2020txg}. Similarly, the GW postmerger spectrum has robust features that can be captured within a few percent accuracy by tidal parameters and/or other properties of NS equilibria in EOS-insensitive way~\cite{Bauswein:2011tp,Hotokezaka:2013iia,Takami:2014zpa,Bernuzzi:2015rla,Breschi:2019srl,Breschi:2021xrx}. These relations have some practical use in GW astronomy because they deliver accurate estimates for the peak luminosity \cite{Zappa:2017xba,Bernuzzi:2020tgt} and for the remnant properties \cite{Bauswein:2012ya,Bauswein:2014qla,Agathos:2019sah} (see also \cite{Bernuzzi:2020tgt} for a detailed review) and because they are the building blocks to develop NR-informed waveform models.

First, we consider the mass-rescaled GW amplitude and frequency at the
moment of merger, $A^{\rm mrg}_{22}/M$ and $Mf^{\rm mrg}_{22}/\nu$, and update the fits developed in Ref.~\cite{Bernuzzi:2014kca,Breschi:2019srl,Breschi:2022xnc}. 
Following closely the fitting procedure of Ref.~\cite{Breschi:2022xnc}, we represent any quantity by the factorized function
\be\label{eq:BreschiFit}
Q^{\rm fit} = a_0 Q^M(X) Q^S(\hat{S},X) Q^T(\kappa^T_2,X)\,,
\ee
where each factor $Q^M$, $Q^S$, $Q^T$ accounts for the mass ratio in terms of $X=1-4\nu$, spin corrections in terms of $\hat{S}$, and tidal effects in terms of $\kappa^T_2$. 
The first two factors are given by the linear polynomial expressions $Q^M = 1 + a^M_1X$, and $Q^S = 1 + p^S_1\hat{S}$, with $p^S_1 = a^S_1(1 + b^S_1X)$. The last factor is instead a rational polynomial
\be
Q^T = \frac{1 + p^T_1\kappa^T_2 + p^T_2{\kappa^T_2}^2}{1 + p^T_3\kappa^T_2 + p^T_4{\kappa^T_2}^2}\,,
\ee
with $p^T_i = a^T_i( 1 + b^T_iX )$. The best fit parameters are shown in Tab.~\ref{tab:fits}. The amplitude and frequency have $1\sigma$ errors of $2.6\%$ and $4.6\%$ respectively. We also obtain a $\chi^2$ of $\sim0.126$ for the former and $\sim0.329$ for the latter.

\begin{figure}[t]
   \centering 
     \includegraphics[width=0.7\textwidth]{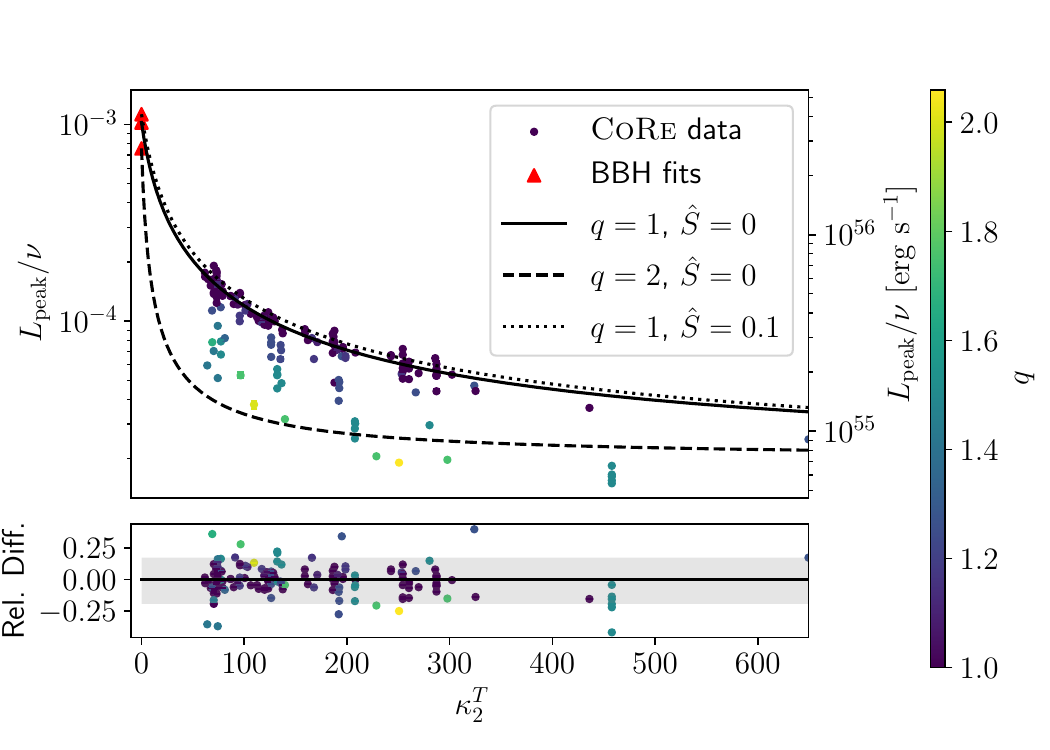}
     \caption{Luminosity peak data from the \core~database. Black lines show the new fits developed for different mass ratio and spin configurations. The relative differences are shown in the bottom panel, where the gray shaded region marks the $90\%$ credible region.
     }
  \label{fig:lpeak}
\end{figure}

Next, we use the public \core~data on the emitted GW energy and
extract the peak luminosity $L_{\rm peak}$ using
Eq.~(\ref{eq:lpeak}). For binary black holes, this quantity does not
depend on the mass scale and it is accurately described by the fits of
Ref.~\cite{Keitel:2016krm}. For BNS, it has been
studied in Ref.~\cite{Zappa:2017xba}. 
We propose the ansatz
\be
L_{\rm peak}(\nu,\,\hat{S},\,\kappa^T_2)/\nu = L^{\rm BBH}_{\rm peak}\frac{1 + p_1(\nu,\,\hat{S})\kappa^T_2 + p_2(\nu,\,\hat{S}){\kappa^T_2}^2}{( 1 + [p_3(\nu,\,\hat{S})]^2\kappa^T_2 )^2}\,,
\ee
where $L^{\rm BBH}_{\rm peak}$ are the mass and spin dependent fits from Ref.~\cite{Keitel:2016krm} and
\begin{eqnarray}
p_k(\nu,\,\hat{S}) &= p_{k1}(\hat{S})\nu + p_{k2}(\hat{S})\nu^2 + p_{k3}(\hat{S})\nu^3 \non\\
p_{kj}(\hat{S}) &= p_{kj0}\hat{S} + p_{kj1}\,.\non
\end{eqnarray}
Note the scaling factor $1/\nu$ for $L_{\rm peak}$. By construction, the fit reduces to the BBH case for $\kappa^T_2\rightarrow 0$.
The luminosity peak is calculated in geometric units; the conversion factor to CGS units is given by the Planck luminosity $L_P=c^5/G \approx3.63\times 10^{59}~\rm{erg}~\rm{s}^{-1}$.
Figure~\ref{fig:lpeak} shows the best fit for $L_{\rm peak}/\nu$ and the
\core~data; the best fitting coefficients are reported in
Tab.~\ref{tab:fits_lp}. The average $1\sigma$ deviation is about
$12\%$ over the entire dataset with less than a dozen of
outliers. The peak luminosities for $q\sim2$ BNS are the least
accurately modelled (4 BNS configurations). The figure shows that the
largest peak luminosities are reached by BNS with $\kt\lesssim80$ that
correspond to high-mass binaries and prompt collapse mergers. These
events can reach peak luminosities of
${\sim}10^{55}~\rm{erg}~\rm{s}^{-1}$, about an order of magnitude less
than binary black holes (of any mass). BNS mass ratios $q\gtrsim1.5$
can lower $L_{\rm peak}$ of about an order of magnitude, while spins
of magnitude ${\sim}0.1$ do not significantly affect $L_{\rm
  peak}$. We stress that BNS with the largest peak luminosity do not
correspond in general to the BNS that radiate the largest amount of
energy because postmerger emission can radiate further energy
\cite{Bernuzzi:2015opx,Zappa:2017xba} (see also Fig.~\ref{fig:ej}). 
We can set an upper limit to the total radiated GW energy from our dataset, obtaining 
$E^{\rm tot}_{\rm GW}\lesssim 0.676~\Msun c^2$.

\begin{figure}[t]
   \centering
     \includegraphics[width=0.7\textwidth]{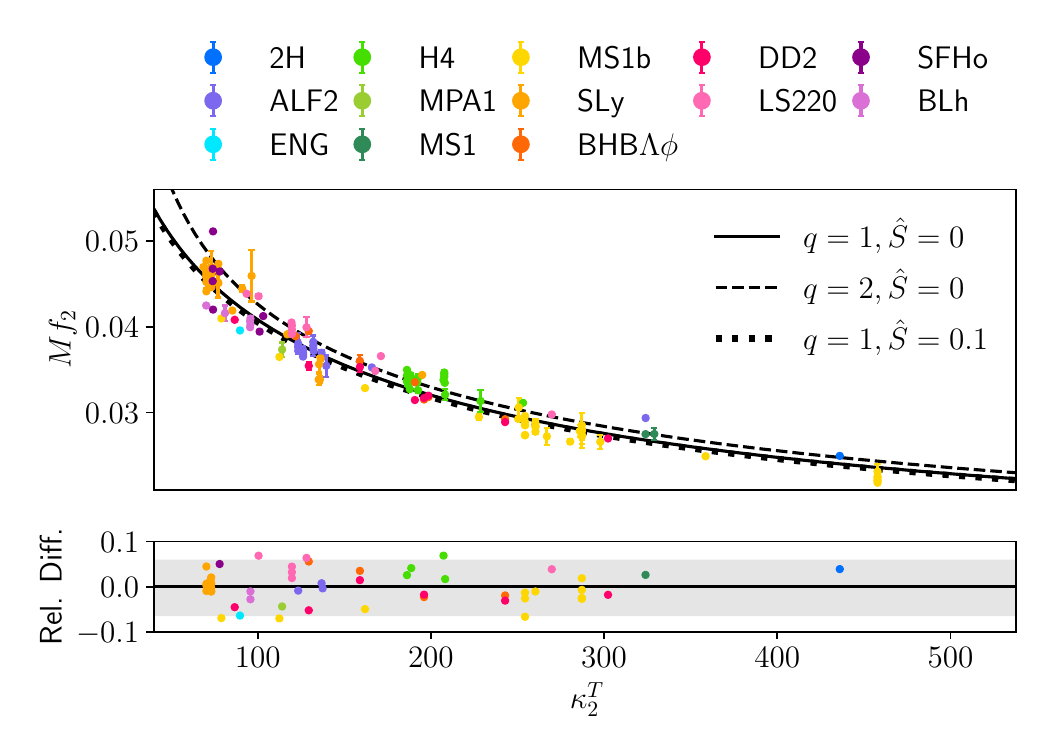}
     \caption{Quasi-universal relation of the post-merger peak frequency $Mf_2$ (mass rescaled) as a function of the tidal polarizability $\kappa^T_2$. Each point represents a simulation of the \core~database with its corresponding EOS (in different colors). Black lines represent the updated $Mf_2$-fits (top panel). The relative differences are shown in the bottom panel, where the gray shaded region marks the $90\%$ credible region.}
  \label{fig:Mf_kap}
\end{figure}

Finally, we illustrate the use of \core~data to model postmerger GWs by discussing a fit of the postmerger's spectrum peak frequency $f_2$, e.g.~\cite{Bauswein:2015yca,Bernuzzi:2015rla,Breschi:2019srl,Breschi:2022xnc}. This peak frequency is a robust feature found in all NR simulations. Direct GW inference on $f_2$ can be used to constrain NS properties \cite{Bauswein:2012ya,Takami:2014zpa,
	Bauswein:2014qla,Chatziioannou:2017ixj,
	Easter:2020ifj,Tsang:2019esi,Breschi:2021xrx}.
The peak frequency also enters as one of the central parameters in
postmerger waveform models that will be employed in the future for more sophisticated matched-filter analyses \cite{Breschi:2022ens}.
Following Ref.~\cite{Breschi:2022xnc} we employ again
Eq.~(\ref{eq:BreschiFit}) to fit the mass-rescaled $Mf_2$. The best
fitting coefficients are presented in Table~\ref{tab:fits} and have a
$\chi^2\sim0.07$. Figure~\ref{fig:Mf_kap} shows $Mf_2$ as a function
of $\kt$ for selected values of mass ratio and spin. 
The $1\sigma$ error is below $4\%$; this precision is
in principle sufficient for 
informative measurements of the NS mass-radius sequence. 
For example, using the EOS-insensitive relation between $f_2$ and the maximum density of an
equilibrium non-rotating NS put forward in \cite{Breschi:2021xrx}, it
would be possible to determine the maximum density of an
equilibrium non-rotating NS to ${\sim}15\%$ and the maximum mass
$\Mmax$ to ${\sim}12\%$ with a single signal at the detectability
threshold.

\begin{figure}[t]
   \centering
     \includegraphics[width=0.7\textwidth]{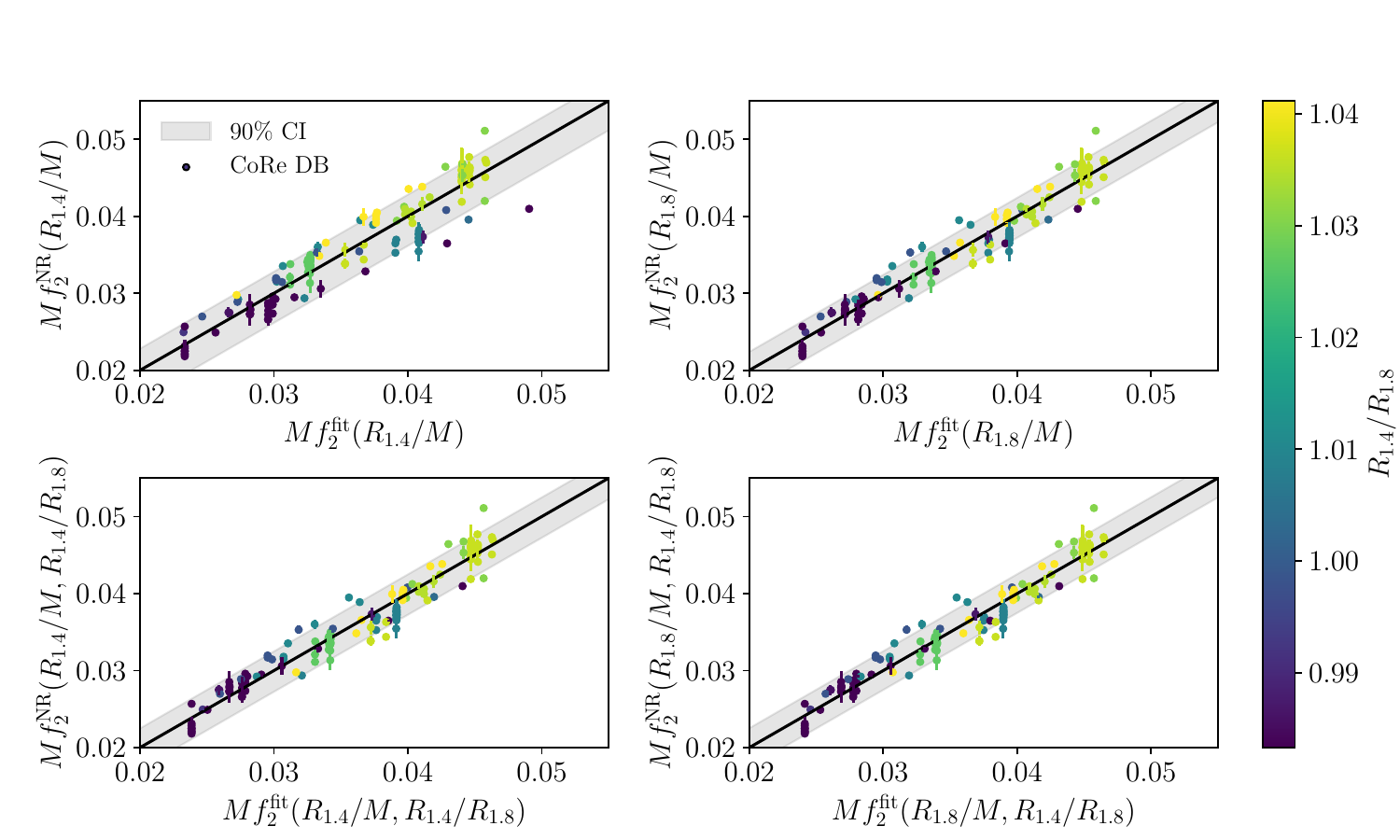}
     \caption{Quasi-universal relation of the post-merger peak frequency $Mf_2$ (mass rescaled) as a function of the NS radius $R_{1.4}$, $R_{1.8}$. Each point represents a simulation of the \core~database with the different colors representing the ratio $R_{1.4}/R_{1.8}$. Each panel shows the different fit calibrations performed in \cite{Breschi:2022xnc}. The black line represents the case when $Mf_2^{\rm{NR}}=Mf_2^{\rm{fit}}$, whereas the gray area shows the $90\%$ credibility level.}
  \label{fig:Mf_R}
\end{figure}

As a further illustration, we calibrate the EOS-insensitive relations (mass-rescaled)
between $f_2$ and the NS radius
\cite{Most:2021ktk,Raithel:2022orm,Breschi:2022xnc}
\begin{eqnarray}
Mf_2\left(\frac{R_X}{M}\right) = a_0 + a_1\frac{R_X}{M} +a_2\left(\frac{R_X}{M}\right)^2 \non\\
Mf_2\left(\frac{R_X}{M}, \frac{R_{1.4}}{R_{1.8}}\right) = a_0 + a_1\frac{R_X}{M} +a_2\left(\frac{R_X}{M}\right)^2 + a_3\frac{R_{1.4}}{R_{1.8}} \,,
\end{eqnarray}
where $R_X$ is the equilibrium radius corresponding
to a NS with mass $X=1.4,\, 1.8~\Msun$. Figure~\ref{fig:Mf_R}
shows $Mf_2$ as function of $R_{1.4}$, $R_{1.8}$ and
$R_{1.4}/R_{1.8}$. Best fit parameters are given in
Table~\ref{tab:fits_Mf2Rx}. Other features of the postmerger spectrum can be
quantified in a similar way. We release reduced postmerger data and
analysis scripts on the \core~website.

\section{Conclusion}

We presented a new set of BNS simulations for the second release of
the \core~database, expanding it to 254 different binary
configurations covering a wide parameter space. The new data includes
BNS consistent with the GW events GW170817~\cite{Nedora:2020pak} and
GW190425~\cite{Dudi:2018jzn,Camilletti:2022jms}. Simulations were
performed with a large number of EOSs, including several microphysical
models~\cite{Camilletti:2022jms,Nedora:2020pak,Prakash:2021wpz}. Some
simulations include the effects of neutrinos, either through the
leakage scheme~\cite{Galeazzi:2013mia,Neilsen:2014hha,Radice:2016dwd},
or using the M0
approach~\cite{Radice:2016dwd,Radice:2018pdn}. Turbulent viscosity is
included in some models using the GRLES formalism
\cite{Radice:2017zta,Radice:2020ids}. Finally, we include simulations
produced using a new hybrid numerical flux scheme, EFL, that was
introduced in~\cite{Doulis:2022vkx} showing fourth order convergence
and smaller phase errors than previous simulations using WENO schemes in BAM. 

We described in detail the methodology we used to assess the overall
accuracy of the waveforms and presented results for all the
configurations in the database. The \core~database waveform have
typical unfaithfulness of less than $10^{-2}$, some have
unfaithfulness of less than $10^{-4}$, so they are suitable for
precision waveform modelling applications. However, to ensure the
convergence and usability of the simulations, more extensive analysis
is needed. As an example, we showed a full analysis of one of our
 simulations, \texttt{BAM:0066}, which showed a clear 
second order convergence and passed several accuracy tests.

Finally, as a first application of the \core~database, we fitted
phenomenological formulas for the merger amplitude, frequency, and GW
luminosity. These fits are able to model the \core~data with high
accuracy ($<5\%$ for the merger amplitude and frequency and $17\%$ for
the peak luminosity). We also recalibrated various quasi-universal
relations between the post-merger peak frequency and the binary
parameters, again finding deviations from the universal relations of
only a few percent. These were used in \cite{Breschi:2022xnc} to
construct the first complete inspiral, merger, and post-merger
waveform model for BNS.

We release the \core~database to the community with the hope that it will enable future discoveries in GW astronomy. Potential applications include the development of new waveform models, the validation of data analysis pipelines and new numerical relativity codes, and the planning of future GW experiments. In the future, we plan to release new simulation data on a rolling basis, with data releases taking place at the publication time of the corresponding paper.

\ack

AG, MB acknowledge partial support from the Deutsche Forschungsgemeinschaft (DFG)
under Grant No. 406116891 within the Research Training Group RTG
2522/1.
FZ, MB, SB, GD acknowledge support from the EU H2020 under ERC Starting
Grant, no.~BinGraSp-714626.
DR acknowledges funding from the U.S. Department of Energy, Office of Science, Division of Nuclear Physics under Award Number(s) DE-SC0021177 and from the National Science Foundation under Grants No. PHY-2011725, PHY-2020275, PHY-2116686, and AST-2108467.
SB acknowledges support from the Deutsche Forschungsgemeinschaft, DFG,
project MEMI number BE 6301/2-1.
BB acknowledges support from the Deutsche Forschungsgemeinschaft, DFG,
Grant BR 2176/5-1.
WT acknowledges funding from the National Science Foundation under Grants
PHY-2011729 and PHY-2136036.

Numerical relativity simulations were performed at various
supercomputing centers.
{\scshape ARA}, a resource of Friedrich-Schiller-Universt\"at Jena
supported in part by DFG grants INST 275/334-1 FUGG, INST 275/363-1
FUGG and EU H2020 BinGraSp-714626.
The authors gratefully acknowledge the Gauss Centre for Supercomputing
e.V. (\url{www.gauss-centre.eu}) for funding this project by providing
computing time on the GCS Supercomputer SuperMUC at Leibniz
Supercomputing Centre (\url{www.lrz.de}, Gauss projects pn29ba, pn56zo, pn68wi).
The authors acknowledge the national High Performance Computing Center Stuttgart (HLRS) for
providing access to the supercomputer HPE Apollo Hawk under
the grant numbers GWanalysis/44189 and INTRHYGUE/44215.
The authors gratefully acknowledge the computing time granted by the Resource Allocation Board and provided on the supercomputer Lise and Emmy as part of the NHR infrastructure, where resources were granted through the project bbp00049.
{\scshape Joliot-Curie} at GENCI@CEA (PRACE-ra5202).
{\scshape Marconi} at CINECA (ISCRA-B project HP10BMHFQQ, INF20\_teongrav and INF21\_teongrav allocation).
{\scshape Bridges}, {\scshape Bridges2}, {\scshape Comet}, {\scshape Expanse}, {\scshape Stampede2} (NSF XSEDE allocation TG-PHY160025), NSF/NCSA {\scshape Blue Waters} (NSF AWD-1811236), supercomputers. This research used resources of the National Energy Research Scientific Computing Center, a DOE Office of Science User Facility supported by the Office of Science of the U.S.~Department of Energy under Contract No.~DE-AC02-05CH11231.

The \core~data are hosted on the \texttt{gitlab} server at TPI Jena.
Data postprocessing was performed on the Virgo ``Tullio'' server 
at Torino supported by INFN.

\section*{References}

\bibliographystyle{iopart-num}
\providecommand{\newblock}{}

\appendix

\section{Public \core~Database }
The simulation data discussed in this work is publicly available at
\begin{center}
\url{http://www.computational-relativity.org/gwdb/}
\end{center}
The database metadata are summarized in the repo \texttt{core-database-index}, which contains a \texttt{json} file with the main properties of the available simulations and the different runs. A repository is associated to one distinct physical binary and contains folders for the different runs performed. For each run, we release a complete metadata file and a HDF5 file with the multipolar waveform for both $\psi_{\ell m}$ and $h_{\ell m}$ at different extraction radii and the energetics. 

Access to our private codes and data is possible upon reasonable request. Any use of the simulation data must be done in accordance with the terms of use contained here: \url{http://www.computational-relativity.org/terms/}.

\section{\watpy: Waveform Analysis Tools in Python}
The repository 
\begin{center}
\url{https://git.tpi.uni-jena.de/core/watpy}
\end{center}
provides classes to work with the \core~waveforms and tutorials. It is also available via \texttt{PyPI}
\begin{center}
\url{https://pypi.org/project/core-watpy/}.
\end{center}
The code includes two main modules.
The {\tt coredb} module contains tools to 
download and upload NR simulation data, 
menage the metadata of the simulations,
visualize statistics of the database,
and work with the HDF5 files provided 
in the \core~website.
The {\tt wave} module
provides methods for the visualization and the analysis of 
(multipolar) NR outputs, i.e. Weyl curvature and GW strain.
\watpy~ is compatible with NR files from BAM, Cactus/Einstein Toolkit (WhiskyTHC/FreeTHC) and the \core~database. 

\section{Merger and postmerger fit data}
The data and scripts employed for the development of the fits presented in this work, can be found in 
\begin{center}
  \url{https://doi.org/10.5281/zenodo.7253784}.
\end{center}
Any reuse of the merger and postmerger fit data must be done in accordance with the Creative Commons Attribution 4.0 International license which applies to the data files.

\section{\THC{}}
\THC{} is open source and publicly available At
\begin{center}
  \url{https://bitbucket.org/FreeTHC/workspace/projects/THC}
\end{center}
A tutorial, microphysics tables, and example parfiles are available at
\begin{center}
  \url{http://personal.psu.edu/~dur566/whiskythc.html}
\end{center}

\end{document}